\documentclass[a4paper,11pt]{article}
\usepackage{a4wide,epsfig,graphics,amsthm,amsfonts,amssymb,amsmath,dcolumn,calc,verbatim,rotating,multirow, ulem,  environ,booktabs,verbatim, pdflscape,rotating,bbm}

\usepackage[paper=a4paper,left=1in,right=1in,top=1in,bottom=1in]{geometry}
\usepackage[utf8]{inputenc}
\usepackage{bm} 
\usepackage{lscape}
\usepackage{natbib}
\usepackage{caption}
\usepackage{ragged2e,array}
\usepackage{arydshln}
\usepackage[shortlabels]{enumitem}
\usepackage[resetlabels]{multibib}
\usepackage{soul}
\usepackage{eurosym}

\usepackage{subcaption}
\expandafter\def\csname ver@subfig.sty\endcsname{}

\usepackage{newtxtext,newtxmath} 

\usepackage[onehalfspacing]{setspace}

\PassOptionsToPackage{hyphens}{url}\usepackage[colorlinks=true,linkcolor=black,citecolor=izablue,backref=page]{hyperref} 
\usepackage{graphicx}
\usepackage{subfig}

\usepackage{tabularx}
\usepackage{graphicx}
\usepackage{adjustbox}
\usepackage{longtable}

\newcites{Web}{Non-Academic References}

\def\sym#1{\ifmmode^{#1}\else\(^{#1}\)\fi}

\usepackage{xargs}   
\usepackage[pdftex,dvipsnames]{xcolor}  
\usepackage[colorinlistoftodos,prependcaption,textsize=tiny]{todonotes}

\usepackage{etoolbox}
\newbool{FDcomments}

\newcommand{\bi}{\begin{itemize}}
\newcommand{\ei}{\end{itemize}}

\booltrue{FDcomments} 

\newcommandx{\FD}[2][1=]{%
  \ifbool{FDcomments}{%
    \todo[inline, linecolor=Plum, backgroundcolor=Plum!25, bordercolor=Plum, #1]{#2}%
  }{}%
}\newcommandx{\AlA}[2][1=]{\todo[inline, linecolor=Green,backgroundcolor=Green!25,bordercolor=Green,#1]{#2}} 

\newcommandx{\JN}[2][1=]{%
  \ifbool{FDcomments}{%
    \todo[inline, linecolor=Blue, backgroundcolor=Blue!25, bordercolor=Blue, #1]{#2}%
  }{}%
}

\definecolor{izablue}{RGB}{0,128,172}

\begin{document}

\renewcommand{\baselinestretch}{1.12}
\title{Hired in High Season: Seasonal Labor Demand and Refugee Labor Market Integration\thanks{University of Potsdam, CEPA, BSoE, \texttt{degenhardt@uni-potsdam.de}. \\
I thank Marco Caliendo, Katrin Huber, Louis Klobes, Jan Nimczik, Aiko Schmeißer as well as seminar participants at the University of Potsdam, Berlin Applied Micro Lunch seminar and ESPE 2025 for valuable comments.
 \\
\indent \textbf{Data Availability:} This paper uses confidential data from the Austrian Labor Market Database (AMDB) maintained by the Austrian Labor Market Service (AMS). The data can be obtained by filing a request directly with the AMS (\url{https://arbeitsmarktdatenbank.at}). The author is willing to assist (\url{degenhardt@uni-potsdam.de}). \\
\indent \textbf{Disclosure Statement:} The author declares that he has no relevant or material financial interests that relate to the research described in this paper. \\

}}

\author{Felix Degenhardt}

\date{\vspace{0.5cm} 
Working Paper \\ 
June 2026}

\maketitle

\begin{abstract}
{\fontfamily{ptm}\selectfont\fontsize{12}{14}\selectfont
\setstretch{2.0}
\noindent I examine whether early but temporary access to low-barrier hospitality employment affects refugees' labor market integration. I exploit within-region, within-year variation by combining the quasi-exogenous allocation of refugees to Austrian regions with seasonality in hospitality, where 25\% of refugees first find work. Labor market access during high seasonal demand raises early employment probability by  3 percentage points (9\% of the mean). Employment gains fade after one year, but treated refugees accumulate significantly higher three-year earnings, with no differences in medium-term wages or job quality. However, early hospitality work increases segregation into refugee-typical industries and firms with fewer Austrian coworkers.

\vspace{0.5cm}

\par}
\end{abstract}

\thispagestyle{empty}

\clearpage
\newpage

\setcounter{page}{1}

\section{Introduction}\label{sec:introduction}

The large inflow of refugees to many European countries since 2015 has intensified policy debates regarding optimal strategies for refugee labor market integration. Since then, calls for a quick take-up of jobs have been particularly salient. However, if institutional barriers such as employment bans and slow skill recognition are high, refugees' work opportunities are oftentimes limited to jobs with particularly low entry barriers. This paper examines whether the fast take-up of only temporarily available low-barrier jobs has an impact on the labor market integration of refugees.

The fast take-up of low-barrier jobs may impact refugee labor market integration in different ways: On the one hand, they may serve as a stepping stone to better jobs by providing access to the labor market, and refugees can accumulate country- or sector-specific skills earlier. On the other hand, low-barrier and potentially low-quality jobs may also trap refugees in worse employment relations if the take-up of such jobs crowds out longer job search and investment in skills \citep{arendt2022,arendt2023}. Previous findings highlight the importance of local labor market conditions for refugee employment, such as higher local labor demand \citep{steinmayr}. However, for refugees mainly the local availability of low-barrier jobs is relevant but has not been studied in detail. One exception is \citet{degenhardtnimczik}, who focus on gig jobs as a specific example of low-barrier employment. 

This paper provides causal evidence on whether the early availability of low-barrier work opportunities in the hospitality sector impacts the labor market integration of refugees in Austria, a country that has experienced one of the highest per-capita refugee inflows among high-income countries since 2015 \citep{eurostat}. The question of how low-barrier employment affects integration trajectories proves particularly salient in Austria, as institutional barriers are high and effectively exclude asylum seekers from employment for long periods. Refugees' recognized educational level in Austria is relatively low, with around 70\% having only primary education. The hospitality sector provides an important entry point to the labor market for refugees who finally gain access to the labor market, comprising approximately 25\% of refugees' initial employment in comparison to 5\% nationally. At the same time, the hospitality sector is highly seasonal, driven by winter (ski tourism) and summer tourism activities. As a result, employment varies substantially within a year, with peaks between June and August and again between December and March. This is an ideal setting to analyze the effects of initial availability of low-barrier jobs, given that refugees who are placed within the same year and local labor market face different opportunities to work in low-barrier jobs.

Using administrative matched employer-employee data from the Austrian Labor Market Database and vacancy data from the Austrian Public Employment Service, I compare outcomes for refugees gaining access to the same local labor market in the same year but facing different labor demand due to seasonal timing. I define the treatment group as those refugees who gain labor market access at the time of high labor demand in the hospitality sector. I compare them with refugees who enter the same region in the same year but in a month with lower labor demand in this sector. Specifically, I use the vacancy-to-unemployment ratio for hospitality jobs at the time individuals gain labor market access as my treatment variable, while controlling for vacancies in all other sectors that may simultaneously affect employment outcomes. A key assumption is that the labor market trajectories of treated refugees would have evolved in parallel as those of the control group. I show evidence that neither the number nor the characteristics of refugees receiving labor market access correlate with my treatment variable, and in particular that refugee admissions show no seasonal patterns that mirror the hospitality sector's seasonal cycle. Specifically, I estimate that a 1\% increase in my treatment variable is associated with only a 0.01\% increase in the number of refugees receiving labor market access, which is statistically insignificant. My design thus allows me to isolate the causal effect of an increase in seasonal labor demand in the hospitality sector from persistent differences in regional economic conditions. To further strengthen my identification strategy, I instrument my treatment variable with pre-refugee-inflow seasonality in the hospitality sector, providing additional evidence that refugee placement and seasonal labor demand are not endogenous to each other. The interpretation of my results remains unchanged.

My identification strategy exploits a unique setting arising from the intersection of Austria's refugee allocation system and pronounced seasonality in hospitality labor demand. Refugees are quasi-randomly assigned across regions and excluded from employment while simultaneously being restricted from relocating until protection status is granted, a process averaging around two years with substantial uncertainty in duration length. This creates quasi-exogenous variation in the time and location of labor market access. This institutional peculiarity, combined with within-region, within-year variation in hospitality labor demand generates plausibly exogenous variation in low-barrier job availability at the time when refugees gain asylum and hence labor market access. 

I consider those refugees who enter in the months of higher  job availability in the hospitality sector as the treatment group and those who enter in other months as the control group and present five main empirical results:

First, treated refugees experience significantly higher initial employment rates in the first year. Specifically, a one standard deviation increase in the treatment variable increases the likelihood to be employed in the first year by up to 3 percentage points, or 9\% of the mean. Effects are largely driven by differences in employment in the hospitality sector. 

Second, employment rates start converging towards the end of the season, as treated refugees are more likely to transition to temporary work agencies and in facility management, while the control group catches up. After around 18 months, employment rates fully converge. Due to higher cumulated employment, treated refugees still have higher cumulated earnings after three years. 

Third, average wages and job quality measures are not significantly different in the first three years after labor market access. The treatment however intensifies labor market segregation, as the likelihood to work in the hospitality sector and sectors more typical for refugees remain higher for treated individuals throughout the observation period. Treated individuals are also more likely to work in firms lower Austrian coworker shares.

Fourth, geographic mobility is largely unaffected, with no significant differences between treatment and control group. However, treated refugees seem to delay their relocation towards the end of a season. This delayed relocation does not translate into better medium-term labor market outcomes among those who eventually move.

Fifth, effects on employment, cumulated employment, and earnings are more positive among refugees with prolonged asylum procedures. Since prolonged asylum processes have been shown to negatively affect employment prospects \citep{hainmueller2016}, these results highlight the importance of early job availability for particularly vulnerable groups. At the same time, the labor market segregation effects are not confined to these groups but persist among refugees with higher predicted employment probability, suggesting that early hospitality work shapes sectoral trajectories regardless of underlying labor market potential.

Overall, my results offer a nuanced perspective on refugee labor market integration through the early take-up of (low-barrier) work. It significantly accelerates the initial employment probability of refugees, and although employment rates converge, cumulated employment and total earnings are significantly higher for treated refugees. Due to the absence of differences in wages and job quality, the conclusion is overall positive and strengthens calls for earlier job take-ups for refugees. Policymakers should however also consider the potential effects on labor market segregation. Refugees react strongly to early low-barrier work opportunities and are likely to stay in the same sector or sectors more typical to refugees even years later, and work in firms with lower Austrian coworker shares, regardless of predicted labor market potential. 

\bigskip

My work relates to the literature examining determinants of refugee labor market integration, with particular emphasis on the role of initial employment conditions in shaping labor market trajectories (see \citealp[]{brell2020labor} for a comprehensive review). This literature identifies multiple factors influencing integration outcomes, including local labor market characteristics such as attitudes against migrants, local unemployment rates and labor market performance \citep{aaslund2007and, aksoy2023first, steinmayr} and local regulations \citep{fasani2021lift,ahrens2024,brunner2014impact,brucker2021occupational}. 

My study contributes to this literature by estimating the effects of early but temporary job availability on refugee labor market integration. Closely related is \citet{steinmayr}, who demonstrate positive employment effects for refugees receiving labor market access in Austrian regions with higher overall job availability, including seasonal employment. My identification strategy goes further by exploiting within-region, within-year variation in hospitality labor demand, which absorbs persistent regional differences in labor market conditions and isolates the effect of short-term, sector-specific job availability from broader regional economic characteristics. I also contribute to emerging literature on how low-barrier employment opportunities affect the labor market integration of refugees. Recent evidence suggests that first work policies and gig economy opportunities not necessarily lead to better labor market integration but potentially crowds out human capital investments \citep{arendt2022,arendt2023,degenhardtnimczik}.\footnote{More broadly, this literature also connects to the literature investigating the impact of first-work policies for the unemployed (see, e.g. \citealp{card2018works}). }

I extend this literature in two dimensions: First, my identification strategy exploits within-year, within-region variation, which absorbs any remaining differences in labor markets not captured by location fixed effects and effectively nets out other potential confounders. Second, I focus exclusively on seasonal variation in the hospitality sector. With approximately 25\% of refugees' employment in their first year concentrated in the hospitality sector in Austria, and similar patterns documented in Switzerland \citep{ahrens2024} and Germany \citep{hauptmann2022beschaeftigung}, this sector represents the most relevant entry point for refugee labor market integration across many European countries. 

I thus complement previous work by focusing on a more traditional type of low-barrier employment. My findings are consistent with the notion that the integration effects of low-barrier employment depend on the job quality gap between early entry jobs and the counterfactual. In Austria, where the overall refugee labor market is concentrated in potentially lower-barrier sectors, hospitality jobs may differ little in quality from the jobs refugees would otherwise take. This contrasts with the gig economy setting of \citet{degenhardtnimczik}, where the quality gap is larger and crowding-out of human capital investment is more plausible \citep{adermon2022gig}. The absence of both lasting employment benefits and human capital distortions in my setting is consistent with this interpretation.

Finally, the study relates to the literature on location choice and geographic mobility of refugees. Recent literature has focused on the influence of local conditions on the location choice of refugees \citep{agersnap2020welfare,dustmann2024refugee,steinmayr} and migrants in general \citep{albert2022immigration}, finding that both the level of welfare benefits and local labor market conditions play a role in the decision to move away from a location. I provide evidence that initial but temporary employment opportunities do not affect whether refugees relocate after receiving freedom to move.  The delayed relocation patterns among refugees exposed to seasonal work, however, suggest that employment duration may influence mobility timing, a finding consistent with evidence on employment-location persistence \citep{steinmayr}. Among the individuals who eventually relocate, likely after the season, I find no significant benefits in medium term career outcomes that would be driven by the higher initial job finding and financial resources.

\section{Labor market and refugees in Austria}\label{sec:institutional_setting}

Austria has received a large number of refugees in recent years, particularly following the large inflows beginning in 2015. Between 2014 and 2019 around 120,000 individuals were granted asylum or subsidiary protection, at times representing up to 1\% of the resident population \citep{bmi, eurostat, steinmayr}.

\paragraph{Institutional setting}
Upon arrival, asylum seekers are required to register at one of three federally operated reception centers. They are then assigned to one of Austria's nine federal states and accommodated in regional asylum shelters within the state. This assignment to the state follows a federal quota system based on population and infrastructure capacity (\textit{Grundversorgungsvereinbarung Art.,15a B-VG}), and is coordinated by federal and state authorities \citep{limberger2010}. Assignment to the shelter is based on infrastructure capacity. Importantly, assignment does not consider integration prospects or local labor market conditions and refugees have no influence over the location of their assignment.\footnote{One exception are family reunions. In my analyses, I however only include adult men, who are least likely to be affected by family reunification \citep{egger2021effects}.} 

While awaiting a decision, asylum seekers are effectively excluded from the labor market and receive basic subsistence support (\textit{Grundversorgung}) from the assigned federal state, which includes housing and a small monetary allowance \citep{rosenberger2012welcoming}.\footnote{\label{footnote:exceptions}A narrow set of exceptions allows them to engage in seasonal or harvest work, non-profit community work, or self-employment. Access to seasonal employment requires a work permit, is capped at six months per year, and is only granted if no suitable unemployed Austrian worker is available. Importantly, taking up any form of paid work typically results in the loss of basic subsistence support, including housing and allowances, which substantially reduces the attractiveness of these legal exceptions. As a result, few asylum seekers make use of this option before receiving protection. I show that the validity of my results is not affected by those who do. Under some strict circumstances, i.e. younger asylum seekers in sectors with high labor demand, asylum seekers can also start a vocational training, asylum seekers can also participate in active labor market trainings. I show that the validity of my results is not affected by those who do.} During the waiting period, asylum seekers are not allowed to move across regions and are required to stay in their assigned shelter, meaning that also mobility within a state is limited. These restrictions remain in place until they are granted asylum or subsidiary protection. Refusing the assigned accommodation or relocating without permission results in the loss of housing and basic subsistence support \citep{rosenberger2012welcoming}.

The asylum process is lengthy, on average taking more than two years. Applicants receive no information about the likely timing of their decision and are not informed about the asylum decision before receiving asylum. This creates considerable uncertainty about when individuals will receive labor market access, reflected in a standard deviation of around 16 months. Combined with the quasi-exogenous geographic allocation and mobility restrictions, this creates quasi-random variation in the timing and location of labor market access. Though the asylum process is lengthy, refugees are largely excluded from work, and training possibilities are limited (see above). I show below that dropping those who take up any of the above does not change the interpretation of my results. Consistent with this argument, the assignment of refugees does not correlate with local labor market or local population characteristics but only with population size \citep[see][]{degenhardtnimczik}, also on the local labor market level. 

Once granted asylum, individuals receive full and unrestricted access to the Austrian labor market, equivalent to that of Austrian citizens. They are no longer eligible for basic subsistence support and must leave their shelter accommodation within four months. Refugees who are granted asylum may also access standard social welfare benefits. Individuals granted subsidiary protection receive the same labor market access but more limited social assistance as they can stay in the basic subsistence support system and the asylum shelter.\footnote{\label{footnote:subs}In my analysis, I treat both refugees and individuals with subsidiary protection jointly, since both groups experience identical assignment and access procedures, and my administrative data do not permit a clean separation. For individuals receiving basic subsistence support, the timing of labor market access may be misclassified as they can rely on basic subsistence support longer. I provide robustness checks that this does not drive my results in section~\ref{sec:robustness}.} In both cases, labor market access begins immediately upon a positive decision. At this point, refugees are free to relocate within Austria and to search for employment on equal footing with native jobseekers.

\paragraph{Austrian labor market}

Once receiving labor market access, refugees navigate in a labor market that is in general characterized by strong industrial relations, including a system of vocational training and occupational regulation. Access to many occupations---particularly those requiring formal training, certification, or licensure---is governed by sector-specific regulations, and the recognition of foreign qualifications is cumbersome and often unsuccessful, which disproportionately often affects refugees \citep{jestl2022trajectories}. 

At the same time, Austria maintains more flexible rules for certain types of employment, particularly seasonal work. Tourism and agriculture sectors rely heavily on short-term labor with limited skill requirements and are exempt from some of the more strict occupational requirements. Seasonal employment contracts can be issued for up to six months, and hiring rules allow for easier labor market access for non-EU nationals, including third-country seasonal workers. These laws create a structural environment in which sectors open to seasonal work, such as the hospitality sector, become especially relevant for migrant workers and refugees entering the Austrian labor market for the first time. 

Similar to other countries, e.g. Switzerland \citep{ahrens2024} and Germany \citep{hauptmann2022beschaeftigung}, the hospitality sector is an important entry point for refugees in Austria, with approximately 25\% of their first-year employment concentrated in this sector. The tourism and hospitality sector overall accounts for a significant share of GDP and employment (both around 5-6\%; \citealp{bundesministerium_wirtschaft}). Austria's hospitality industry is geographically concentrated in alpine regions and along major lakes, with employment driven by two distinct seasonal peaks: a winter season centered on ski tourism running roughly from December to March, and a summer season peaking between June and August. This pronounced seasonality creates substantial within-year variation in labor demand. More broadly, refugee employment in Austria is highly concentrated: 75\% is concentrated in only ten two-digit industries, compared to one third of overall Austrian employment in the same industries.

\subsection{Data and summary statistics}

\paragraph{Data Sources} I use administrative data from the Austrian Labor Market Data Base (AMDB), a matched employer-employee database. The data include detailed daily information on employment and unemployment episodes, transitions into and out of different labor market states, and annual earnings since 1997. My regional unit of observation are labor market regions (AMS-regions). Labor market regions in the AMDB are defined as the area of responsibility for each public employment service center. I combine all different centers in Vienna to one regional labor market, which leaves a total of 88 labor market regions. Thus, refugees are assigned to shelters within one of those labor market regions.  For part of my analyses, I use the matched employer-employee data in the AMDB to compute firm and worker fixed effects following the model by \citet[][henceforth AKM]{abowd1999high} and \citet{degenhardtnimczik}. I use the estimated AKM firm fixed effects to proxy for the unobserved quality of firms. A downside of the AMDB data are top-coded wages, which should not play a role in my setting as less than 0.1\% of refugees in my sample have earnings at this bound. Further the AMDB does neither contain information on education or general individual skills nor occupational information.

To measure local labor demand, I supplement the AMDB with vacancy data provided by the Austrian Public Employment Service. These data cover the universe of vacancies posted through the Public Employment Service's job platform, which accounts for roughly 60\% of all job postings in Austria. Importantly, coverage of job postings is more representative for jobs with general or lower skills \citep{mueller2024vacancy}. For my setting, this is beneficial since the restaurant jobs relevant to my research question are primarily lower-skilled jobs. Further, lower-skilled jobs are also more relevant for refugees, as around 70\% have primary education \citep{steinmayr}. Each vacancy record includes the 4-digit occupation, posting date, and labor market region. I use this information to construct monthly measures of seasonal job demand across labor market regions.

The AMDB also allows identification of specific population groups based on the source of social insurance coverage. I use this feature to identify asylum seekers via their enrollment in a dedicated health insurance scheme tied to the basic subsistence support system (\textit{Grundversorgung}), following \citet{degenhardtnimczik}. A positive asylum decision is proxied by the end of this coverage followed by a transition into employment, registered unemployment, or another insured status within the social security system.\footnote{See footnote \ref{footnote:subs} for potential misclassification and robustness checks in  section~\ref{sec:robustness}.} To differentiate between those who had their asylum claim rejected and would have to leave the country, I characterize positive asylum decisions as those transitioning to any other labor market status and who remained observable in the data for at least six months following the last spell in the health insurance for asylum seekers. Refugees who are not observable right after having their last observation in the health insurance for asylum seekers may either not have gotten a positive asylum decision that ended the asylum seeker status but rather had to leave the country (e.g. due to a negative asylum decision), or they have moved to another European country after their positive asylum decision. The numbers of positive asylum decisions in my sample is largely in line with the asylum statistics provided by the Federal Ministry of the Interior (see \citealp{degenhardtnimczik}).

\paragraph{Summary statistics}

I restrict my sample to refugees aged 18 to 55 who received a positive asylum decision and hence labor market access between 2014 and 2019. This time period includes the large inflow of refugees while leaving enough time for their lengthy asylum processes to be finished. I focus on male refugees to minimize potential confounding from family reunification policies, which disproportionately affect women.\footnote{These sample restrictions largely follow \citet{degenhardtnimczik} and \citet{dellinger2021impact}. Restricting to only positive asylum procedures leads to a sample of 95,000 individuals, a number that is smaller but follows similar trends to the asylum statistics provided by the Federal Ministry of the Interior. I additionally exclude refugees with extremely short asylum procedures (shorter than 30 days), those who had labor market positions before their asylum status and those with missing location or nationality. This reduces the sample size to roughly 80,000. Restricting to male refuges reduces the sample size to around 48,000, age restrictions to the final sample size.} The final sample includes over 35,000 individuals, whom I follow for up to three years after labor market access. Because the Covid-19 pandemic heavily influenced the hospitality sector, I limit the observation period to the end of 2019. I show below that my results are not affected by some individuals not being observable for three years. 

Column (1) of Table~\ref{table:summary_stats_refugees} shows socio-demographic characteristics of refugees in my analysis sample at the time of labor market access and descriptive labor market and mobility outcomes one year after. Panel A shows that, at the time of labor market entry, refugees are on average around 29 years old and, and around 70\% come from either Syria or Afghanistan. The asylum process takes on average more than two years, with a standard deviation of around 16 months. Panel B shows that around a quarter of refugees are in some kind of labor market training offered by the Austrian Public Employment Service one year after labor market access. Around 5\% are in vocational training, which is typically a three-year training program with a specific firm and targeted mostly towards younger individuals.  Among the one third of refugees who are employed, more than one quarter are in the hospitality sector, underlining its importance for refugees in Austria. The average cumulated earnings after one year are on average almost \EUR{5000}. As highlighted by \citet{steinmayr}, refugees disproportionately often move to Vienna, which averages to around 60\% in my sample. Most of the moves away from the assigned location take place within the first couple of months after labor market access, with the average duration to the move to Vienna being less than a month. I discuss the implications of these moves for my empirical design below. 

Column (2) shows descriptive statistics for the subsample of refugees who at some point in the observation period work in the hospitality sector. Panel A shows that those individuals are younger and have waited for their asylum decision slightly longer. They are less likely to be in any type of training and more likely to be employed one year after labor market access, and to some degree mechanically so in the hospitality sector. The high share of employment in the hospitality sector in the first year highlights the importance in the immediate period after labor market access, as column (2) conditions on employment in the hospitality sector at some point within the three years after labor market access. While average wages for those individuals are slightly lower in the first year, cumulated earnings are almost 50\% higher. Initial job finding seems to be a little bit faster, while they also take up more new jobs have slightly lower tenure in the first year.

Though these differences illustrate the potential impact that the hospitality sector has on the labor market integration of refugees, it is clearly reflecting selection into such jobs. To estimate the effect of seasonal hospitality job availability on the labor market integration of refugees, I exploit the seasonality in the Austrian hospitality sector combined with the quasi-random allocation of refugees across regions in Austria.

\section{Empirical approach}

The key idea is to compare refugees who receive labor market access in the same year and region. Because they are quasi-randomly placed in one of those regions and wait on average for two years for their asylum decision, refugees do not have influence on the timing and location at the time of the positive asylum decision and hence labor market access. The only difference between treated and control group is whether they receive labor market access in a time of the year with relatively high vs. low demand for hospitality jobs. To illustrate this comparison, I first document the seasonal patterns in hospitality employment and construct a measure of labor demand for hospitality work. 

\begin{figure}[h!]
\begin{center}
\caption{Seasonal variation in Austrian hospitality industry and refugee labor market access}
\label{fig:seasonality}

\begin{subfigure}[b]{0.6\textwidth}
        \centering
\includegraphics[width=\textwidth]{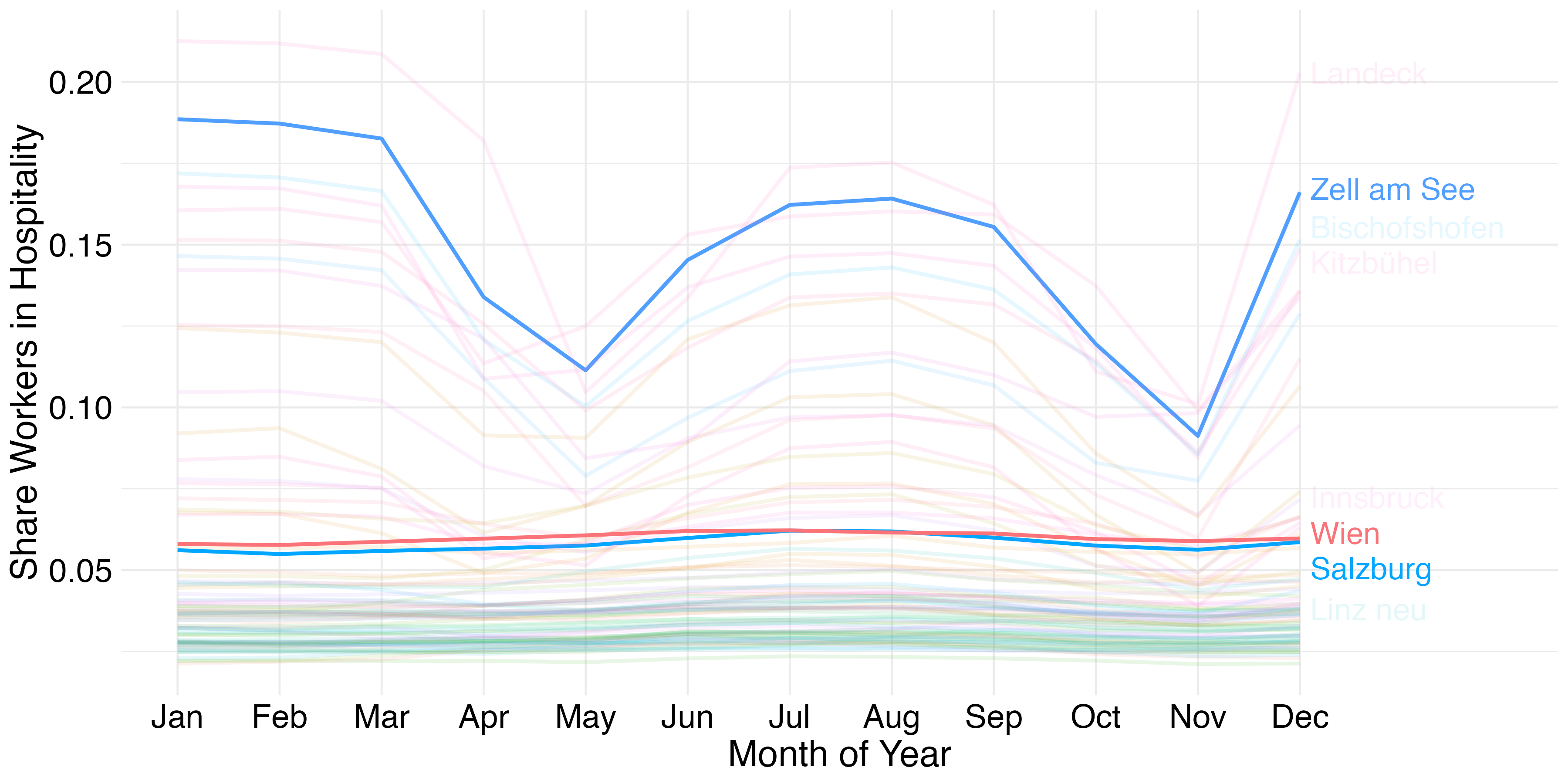}
\subcaption{Within-year variation of employment share}  \label{fig:seasonality_employment}
\end{subfigure}
\begin{subfigure}[b]{0.6\textwidth}
        \centering
\includegraphics[width=\textwidth]{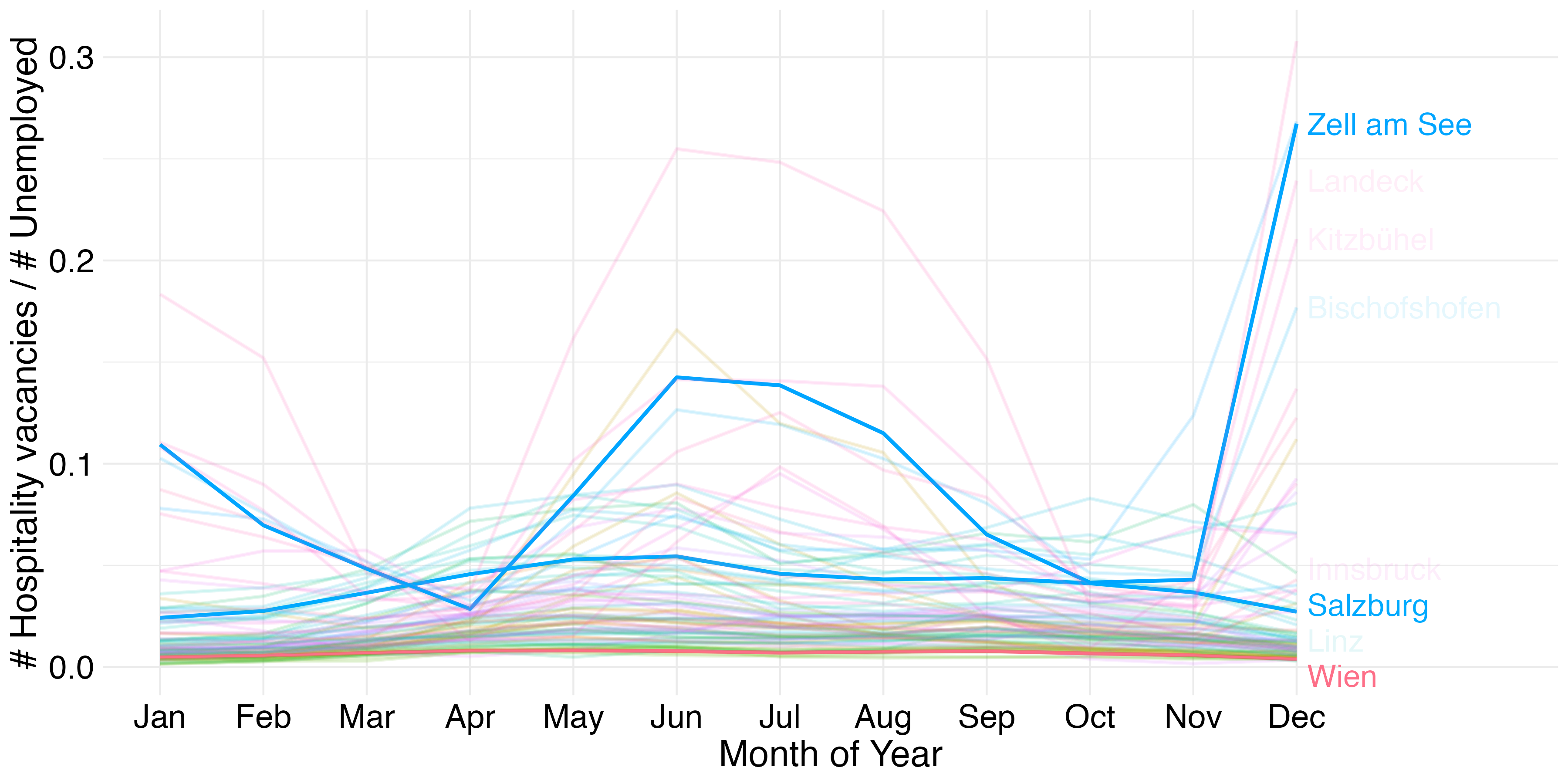}
\subcaption{Within-year variation of vacancy/unemployment ratio} \label{fig:seasonality_vacancies}
\end{subfigure}
\begin{subfigure}[b]{0.6\textwidth}
        \centering
\includegraphics[width=\textwidth]{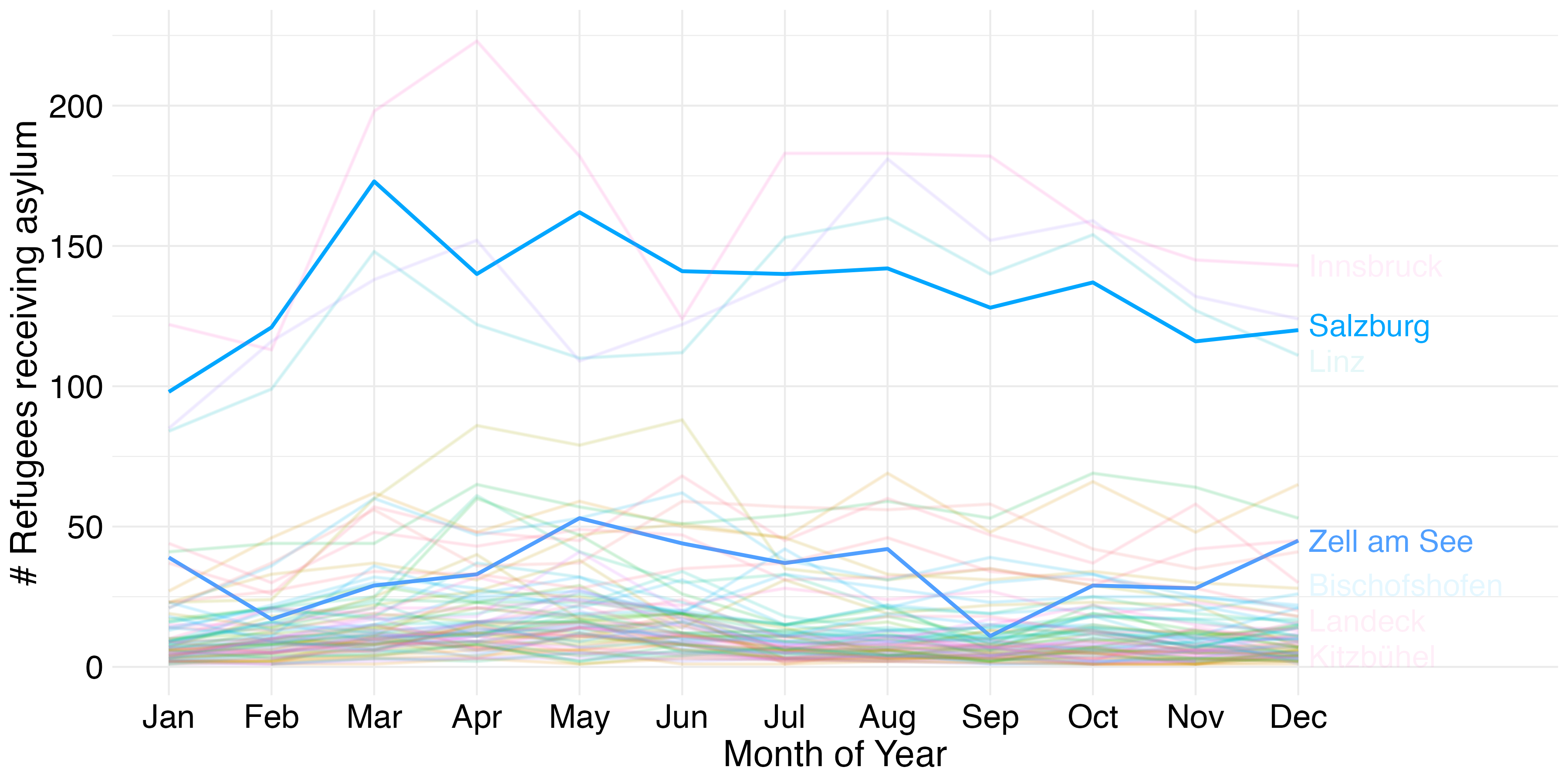}
\subcaption{Within-year variation of the number of refugees who received labor market access} \label{fig:seasonality_refugees}
\end{subfigure}
\end{center}
 \textit{Notes}: This figure shows the seasonality of the Austrian hospitality sector and refugee labor market access. Values for Panels (a) and (c) are based on the AMDB, values for Panel (b) on AMS vacancy data. All values cover the years 2014 to 2019 and are aggregated and averaged by month-of-year and AMS-region, with AMS-regions for Vienna combined into one region. Panel (a) shows the share of workers employed in the hospitality sector in Austria relative to overall employment. Panel (b) shows the number of vacancies for hospitality jobs posted through the Austrian Public Employment Service's platform, divided by the number of unemployed. Panel (c) shows the number of refugees who received labor market access, excluding Vienna to improve readability; Figure~\ref{fig:seasonality_refugees_vienna} includes Vienna.

\end{figure}

Figure~\ref{fig:seasonality} Panel (a) shows seasonality in the Austrian hospitality sector, measured as the share of hospitality workers relative to overall employment. The hospitality sector is defined by industries in the two-digit NACE-codes 55 (accommodation) and 56 (food and beverage service activities). The hospitality sector is a very important part of the overall workforce, particularly so for some labor market regions, many of which are located in alpine areas. Employment shares reach almost 25\% in some months of the year, with very apparent seasonality. In other areas, such as Vienna, the hospitality sector makes around 5\% of employment with substantially weaker seasonal variation. As described above, a lot of refugees move to Vienna immediately after receiving labor market access. Since I compare refugees based on the labor demand for hospitality jobs when receiving labor market access, a potential source of bias for the labor market outcomes would come from differences in moving patterns due to hospitality job availability. I show below that this, however, does not affect the interpretation of my results.

To measure the labor demand in the hospitality sector, I use data from the Austrian Public Employment Service and construct a measure that reflects the vacancies for hospitality jobs to unemployment ratio (HVUR):\footnote{Specifically, these jobs and Austria-specific occupation codes are: waiters/waitresses (B5121), restaurant cooks (B5210), kitchen assistants (B5251), buffet and bar staff (B5131).}
	\begin{equation*}
    \text{HVUR}_{l\mathbf{m}(y)} = \frac{\text{\# Hospitality Job Vacancies}_{l\mathbf{m}(y)}}{ \text{\# All Unemployed}_{l\mathbf{m}(y)}}.
\end{equation*}

This measure is constructed for every labor market region and month of the year cell in my observation period from 2014 to 2019, meaning that it reflects within-year seasonality, not year specific shocks.

Figure~\ref{fig:seasonality} Panel (b) shows the $HVUR$ for each month of the year and labor market region, reflecting strong seasonality within each year. The seasonality in labor demand align with those shown for the employment shares. At the beginning of a season, e.g. in June and December in Zell am See, demand for hospitality jobs is very high. While employment shares remain high during the season, demand for hospitality jobs subsequently decreases, likely reflecting job filling. 

I use this vacancy-to-unemployment ratio $HVUR$ at the time of labor market access as my treatment variable, standardized to have mean zero and variance of one to simplify interpretation. A key assumption is that the labor market trajectories of treated refugees, who receive labor market access during times of higher seasonal labor demand, would have evolved in parallel as those receiving access during times of lower seasonal labor demand, i.e. the control group. One potential concern is strategic granting of positive asylum decisions if seasonal demand for hospitality jobs is high, as refugees are an important part of the hospitality workforce. Figure~\ref{fig:seasonality} Panel (c) shows that the average number of refugees for each month of the year and labor market region does not show seasonal patterns similar to the hospitality labor demand. 

To address this concern even more specifically, I regress the log number of refugees who receive labor market access in each month-labor market region pair on the log of my treatment variable. Column 1 of Table~\ref{tab:validating_treatment_number} shows that there is no significant correlation between the number of refugees receiving labor market access and the treatment variable, with an estimated elasticity of 0.01. Columns 2-6 show that there is also no significant correlation when regressing the level of the HVUR on the number of refugees and average refugee characteristics at the time of labor market access. If younger individuals or individuals from certain countries of origin are more likely to work in the hospitality sector, there may be strategic granting to these groups. As columns 3-6 of Table~\ref{tab:validating_treatment_number} show, I find no evidence for this. Columns 7-10 show evidence that treated refugees are not more likely to have used one of the exceptions to be employed, employed in the hospitality sector, self-employed before labor market access, or to be in labor market training before receiving labor market access. To further strengthen my identification strategy, I instrument my treatment variable with pre-refugee-inflow seasonality in the hospitality sector, providing additional evidence that refugee placement and seasonal labor demand are not endogenous to each other in section~\ref{sec:robustness}. The interpretation of my results remains unchanged.

These results also provide evidence that support for the absence of another important potential threat to validity. As described in section~\ref{sec:institutional_setting}, I can only proxy the timing of granted asylum and hence labor market access in the data by the end of the health insurance for refugees, which is tied to the basic subsistence support. Some refugees may prolong their dependence on basic subsistence support, particularly so those who are granted subsidiary protection. Among Afghans, the share of subsidiary protection is highest (see \citealp{steinmayr}). If these refugees prolong their dependence of basic subsistence support until the season starts, this would threaten the validity of my results. As there is no correlation between the treatment variable and the number of refugees or the country of origin, I interpret this as evidence in support of my proxy for the positive asylum decision.

\subsection{Estimation strategy}
For my estimation strategy, I consider those refugees who enter in the months of higher availability of hospitality jobs as the treatment group and those who enter in other months as the control group. To estimate the causal effect of the treatment on the employment trajectories of refugees, I estimate the following event-study-style specification:\footnote{My empirical approach is methodologically similar to e.g. \citet{oreopoulos2012short} who estimate the career effects of graduating in a recession.}

\begin{align}\label{eq:event_study_dynamic}
\begin{split}
    Y_{i,t} &= \alpha_{l\mathbf{y}(i,t^0)} + \lambda_{m} + \beta_{t} \text{HVUR}_{l\mathbf{m}(i,t^0)}  + \Gamma_t X_{il\mathbf{m}(i,t^0)}  + \epsilon_{il,t},
\end{split}
\end{align}

where the outcome variable $Y_{i,t}$ measures the labor market status of individual $i$ at time $t=[0,36]$ months after gaining access to the labor market in location $l$. $t=0$ is the period of receiving labor market access. Outcomes include employment indicators, wages, job quality measures (e.g., tenure or coworker composition), participation in training, and mobility outcomes. For the dynamic part of the analysis, I estimate equation~\eqref{eq:event_study_dynamic} for each time period $t$ individually. The main coefficient of interest is $\beta_{t}$, which captures the effect of a one standard deviation increase in the vacancy-to unemployment ratio for hospitality jobs, $\text{HVUR}$, for individual $i$ receiving labor market access in location $l$ in month of the year $m$. This compares, for example, receiving labor market access in Zell am See in June vs August, or in Salzburg between February and June. 

The specification includes location-by-year-of-entry fixed effects, $\alpha_{l\mathbf{y}(i,t^0)}$, which flexibly absorb any time-invariant differences across regions as well as broader regional trends across cohorts. $\lambda_{t}$ are calendar month fixed effects that capture any time trends occuring in all regions simultaneously. The identifying variation therefore comes from within-region and within-year differences in hospitality labor demand at the time of labor market access, i.e., across calendar months within the same year and location. Standard errors are clustered at the interacted labor market region (AMS-region) and year of labor market entry as well as calendar month level.

The vector of controls $X_{il\mathbf{m}(i,t^0)}$ importantly includes the non-hospitality va\-can\-cy-to-un\-em\-ploy\-ment ratio at entry, calculated analogous to the $HVUR$ but for all jobs outside of the hospitality sector. This captures any variation seasonal demand in other sectors or sudden shifts in overall labor demand. It also captures variation in the overall unemployment rate that may confound the treatment variable. Additionally, $X_{il\mathbf{m}(i,t^0)}$ includes age at labor market access, length of the asylum procedure, and 21 regions of origin. 

When aggregating the effects, I use a static specification that pools outcomes over the three-year period after labor market access:

\begin{align}\label{eq:event_study_static}
\begin{split}
    Y_{i} &= \alpha_{l\mathbf{y}(i,t^0)} + \lambda_{t} + \beta \text{HVUR}_{l\mathbf{m}(i,t^0)}  + \Gamma X_{il\mathbf{m}(i,t^0)}  + \epsilon_{il},
\end{split}
\end{align}

where $Y_i$ represents outcomes for the entire post-treatment period, with control variables, fixed effects and standard errors being equal to equation~\eqref{eq:event_study_dynamic}.

\section{Results}
\subsection{Effects on employment and mobility}

\begin{figure}[h]
\begin{center}
\caption{The effect of seasonal hospitality jobs on refugee employment}
\label{fig:results_did}

\begin{subfigure}[b]{0.75\textwidth}
    \centering
    \includegraphics[width= \textwidth]{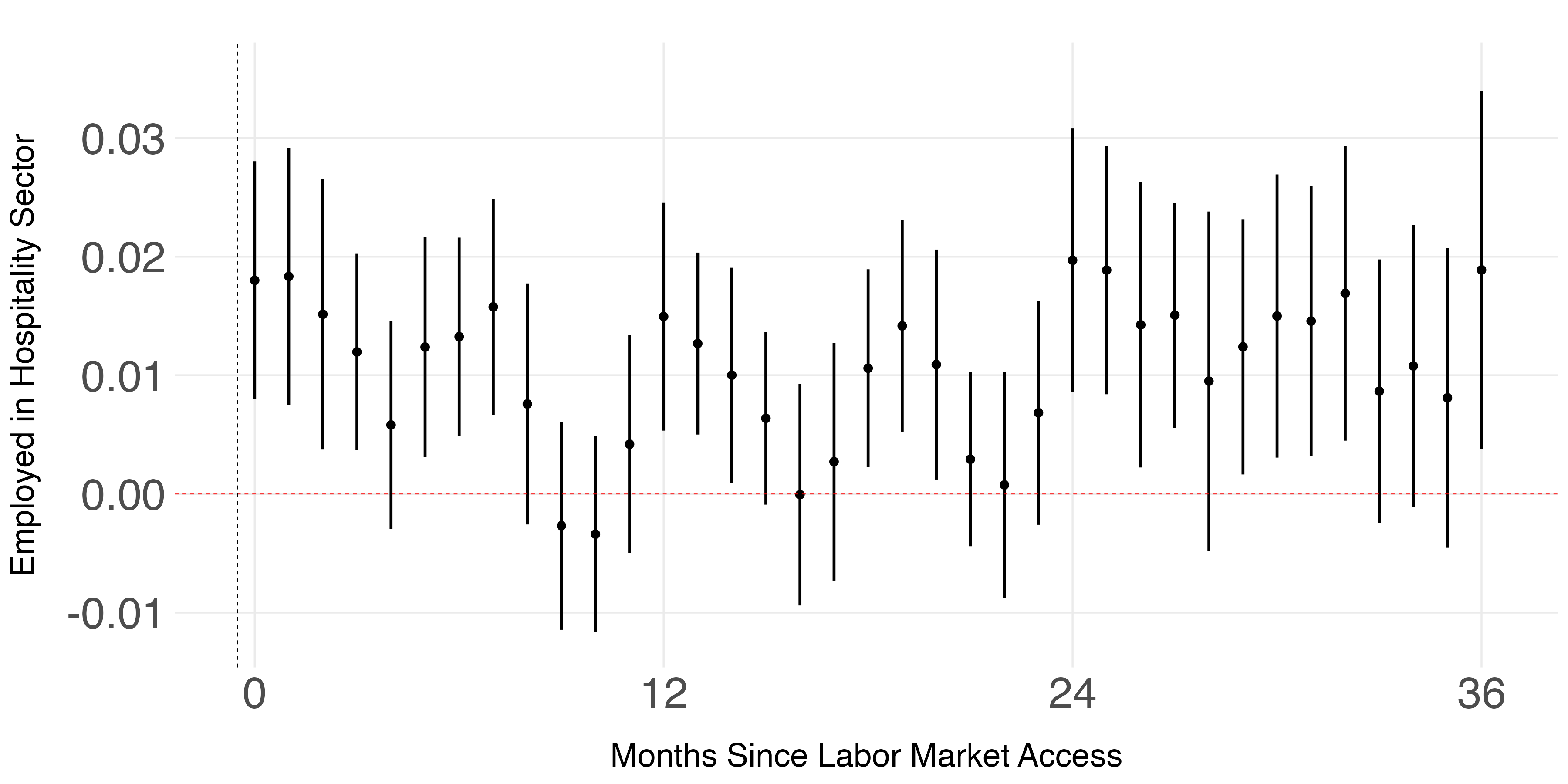}
	\label{fig:results_did1}
\end{subfigure}

\subcaption{Employed in Hospitality Sector}

\begin{subfigure}[b]{0.75\textwidth}
    \centering
\includegraphics[width=\textwidth]{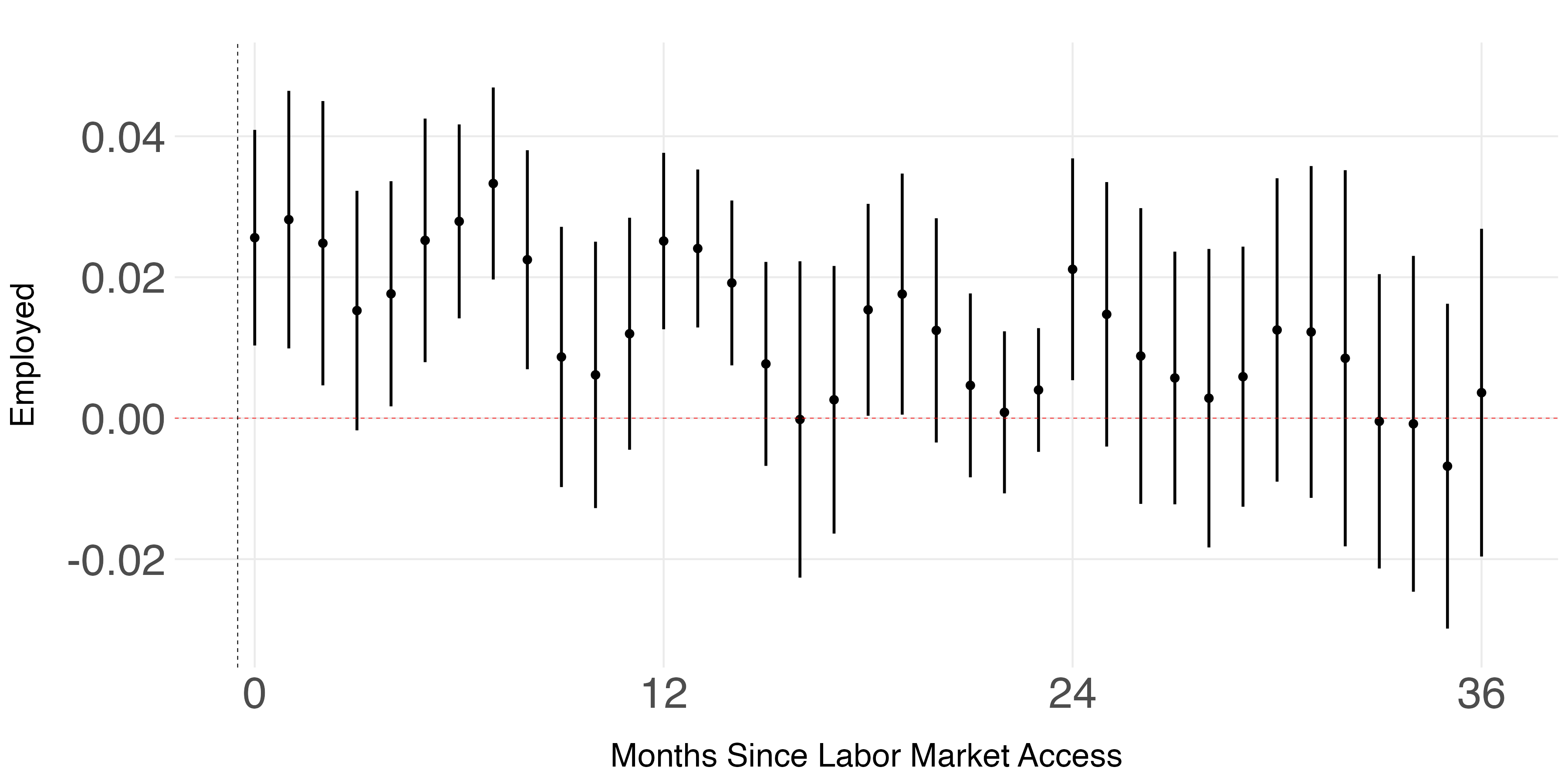}
	\label{fig:results_did2}
\end{subfigure}

\subcaption{Employed}

\end{center}
 \textit{Notes}: This figure shows monthly coefficients for the $HVUR$ from estimating Equation\eqref{eq:event_study_dynamic}) in my sample of refugees. Panel (a) plots the effect on employment in hospitality firms. Panel (b) shows results for the effect on the overall probability to be employed. Bars indicate 95\% confidence intervals. 
\end{figure}

Panel (a) of Figure~\ref{fig:results_did} shows the effect on the likelihood to be employed in the hospitality sector, as estimated by equation~\eqref{eq:event_study_dynamic}. Treated refugees are almost 2 percentage points more likely to work in the hospitality sector in the first three months, which corresponds to around 20\% of the outcome mean. This substantial effect shrinks after the first year but increases again, leading to a persistingly increased likelihood throughout the whole observation period of three years after labor market access, which is characterised by semi-annual seasonal patterns. 

Panel (b) shows that this increase in the likelihood of working in the hospitality sector translates into a higher overall employment probability in the first year, reflecting the same seasonal patterns. The effect on overall employment is also large, with a maximum effect of around 3 percentage points, corresponding to around 9\% of the outcome mean. At the end of the first year, this difference starts to decline and becomes zero towards the end of the observation period. Taken together, one can see that treated refugees seem to have their labor market integration accelerated, while refugees in the control group catch up and resulting employment rates converge. 

Columns 1 and 2 of Table~\ref{table:main_results} show the aggregated employment effects over the whole period of three years after labor market access. On average, a one standard deviation increase in the HVUR at labor market access results in a increase of 1.1 percentage point in the probability to work in the hospitality sector. This average effect is substantial and amounts to over 12\% of the mean in the first three years. The overall employment effects are also positive and significantly different from zero, indicating a 1.4 percentage point increase in the likelihood of being employed, or 4\% of the mean. This means that the aggregate employment effect is positive despite the convergence of employment probability in the third year. Column 3 shows the estimate for employment outside of the hospitality sector, where I subtract those who work in the hospitality sector from the overall employment. The coefficient shrinks towards zero, suggesting that almost the entire aggregate employment effect is driven by the increase in the take-up of hospitality jobs. Column 4 of Table~\ref{table:main_results} additionally shows that treated refugees find a job faster, averaging around 0.3 months earlier for a one standard deviation increase of the treatment variable. These results are striking, given that these differences are driven solely by availability in the initial month of receiving labor market access, while individuals in the control group would face those opportunities merely months later. 

The faster take-up of jobs could affect refugees' mobility decisions. On one hand, consistent with findings from a similar setting in Austria \citep{steinmayr}, higher employment likelihood could increase the probability to stay in the assigned region since jobs typically play an important role in relocation decisions. On the other hand, higher employment likelihood may lead to more financial independence and increase the moving likelihood, especially since refugees have to stay in assigned regions and depend on welfare benefits for extended periods before receiving labor market access. In Austria, refugees disproportionately often move to Vienna, although this is not a particularly strong labor market. I thus follow \citet{steinmayr} and estimate the likelihood to move to Vienna as the main outcome for mobility. 

Column 5 of Table~\ref{table:main_results} shows no significant effects of hospitality job availability on the likelihood to move to Vienna. One reason for that is that a lot of moves take place within the first months after receiving labor market access and hence the freedom to move. The availability of hospitality jobs may not play a larger role for the average likelihood for that decision. However, although the average effects on mobility are zero, I find some evidence that the mobility decision may at least be affected to some degree. Colum 6 shows that for those moving to Vienna within the three years, their move takes place on average half a month later for treated individuals, although this estimate is not significantly different from zero. Table~\ref{table:appendix_results} adds additional evidence for this by estimating equation~\eqref{eq:event_study_static} separately for the months in a labor market region with above and below median $HVUR$ to get an idea whether individuals rather move in high- or low seasons. The results point to the conclusion that the treatment group tends to move to Vienna during low season rather than high season. This offers a possible explanation for the null effect on mobility, as individuals may simply postpone their move until the season ends and adjust the timing of their mobility decision, not their overall mobility decision. 

An important question is whether refugees who delay relocation to work in hospitality first subsequently fare better after moving than those who relocate immediately. Higher initial earnings could in principle facilitate better location choices or investment in job search after relocation. Table~\ref{table:movers} restricts the sample to refugees who eventually relocate and compares treated and control refugees within this group in the medium-run of three years. The results show no significant differences across any outcome as employment, cumulated earnings, wages, job stability, or firm quality are all statistically indistinguishable between treated and control movers, suggesting that the timing of the move relative to the hospitality season does not shape longer-run integration trajectories among those who relocate. Conversely, I show below that excluding those who moved does not change the interpretation of the employment results.

\paragraph{How does the control group catch up?}

With overall employment rates converging in the third year, this is an important question in understanding how the treatment affects the labor market integration of refugees in Austria. I consider two mechanisms: First, whether hospitality job take-up crowds out human capital investment and second, whether it crowds out employment in other, possibly better-paying sectors and firms.

Human capital investments may be crowded out due to the fast take-up of low-barrier work, as suggested by previous research \citep{arendt2022,arendt2023,degenhardtnimczik}. This could then result in refugees being employed in more unstable or lower-paid jobs. Table~\ref{table:appendix_results} shows that aggregate effects on both labor market training and vocational training, two different types of human capital investments, are not significantly different from zero. Although the estimate on labor market training take-up is slightly negative, there is no clear evidence that it crowds out human capital investments.\footnote{Estimates for the first year only are also not significantly different from zero. Results are available upon request.} 

\begin{figure}[h]
\begin{center}
\caption{The effect of seasonal hospitality jobs on refugee employment by sector}
\label{fig:results_did_sectors}

\begin{subfigure}[b]{0.32\textwidth}
    \centering
    \includegraphics[width=\textwidth]{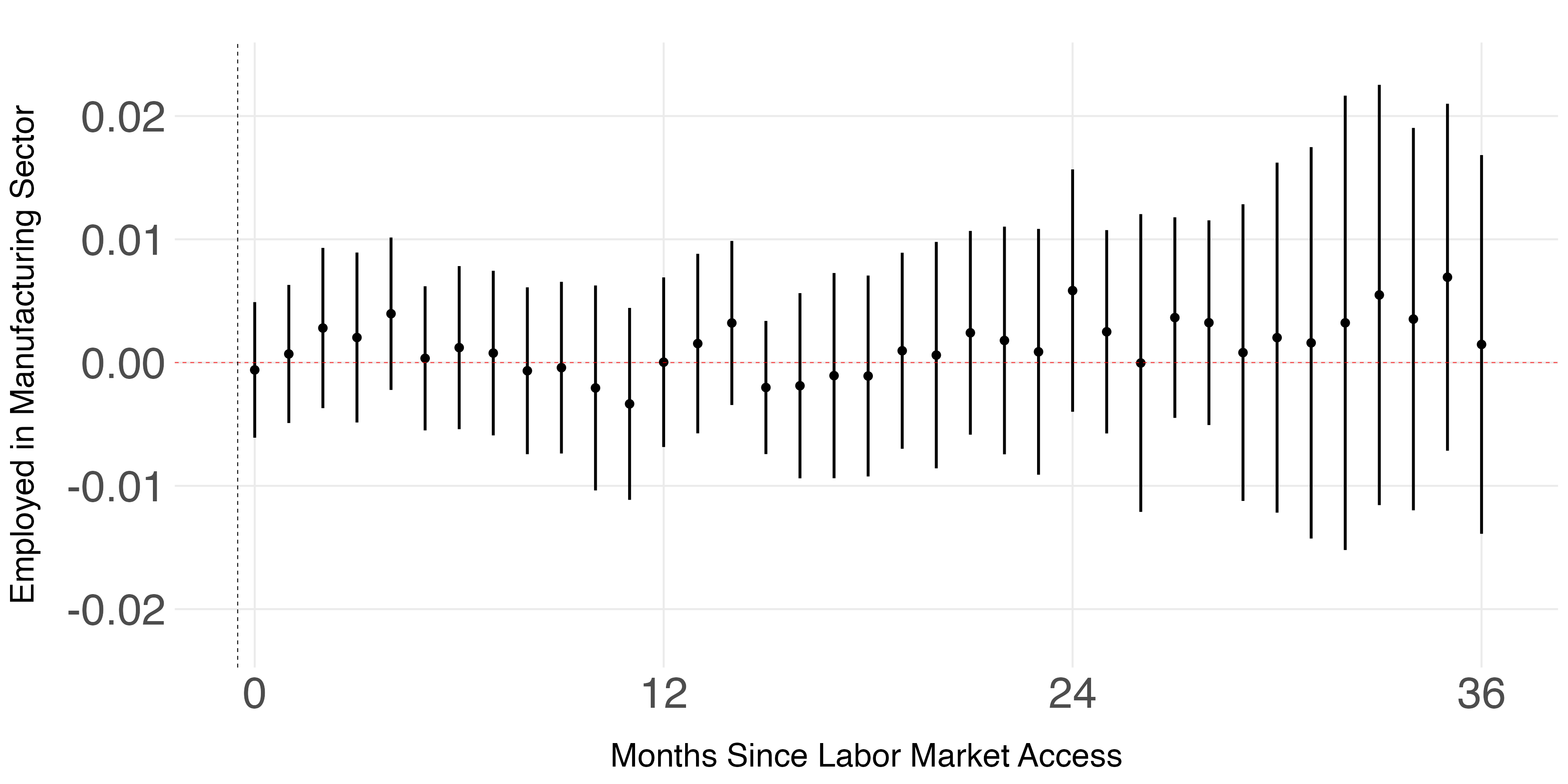}
    \label{fig:results_did_manufacturing}
    \subcaption{Manufacturing Sector}
\end{subfigure}
\hfill
\begin{subfigure}[b]{0.32\textwidth}
    \centering
    \includegraphics[width=\textwidth]{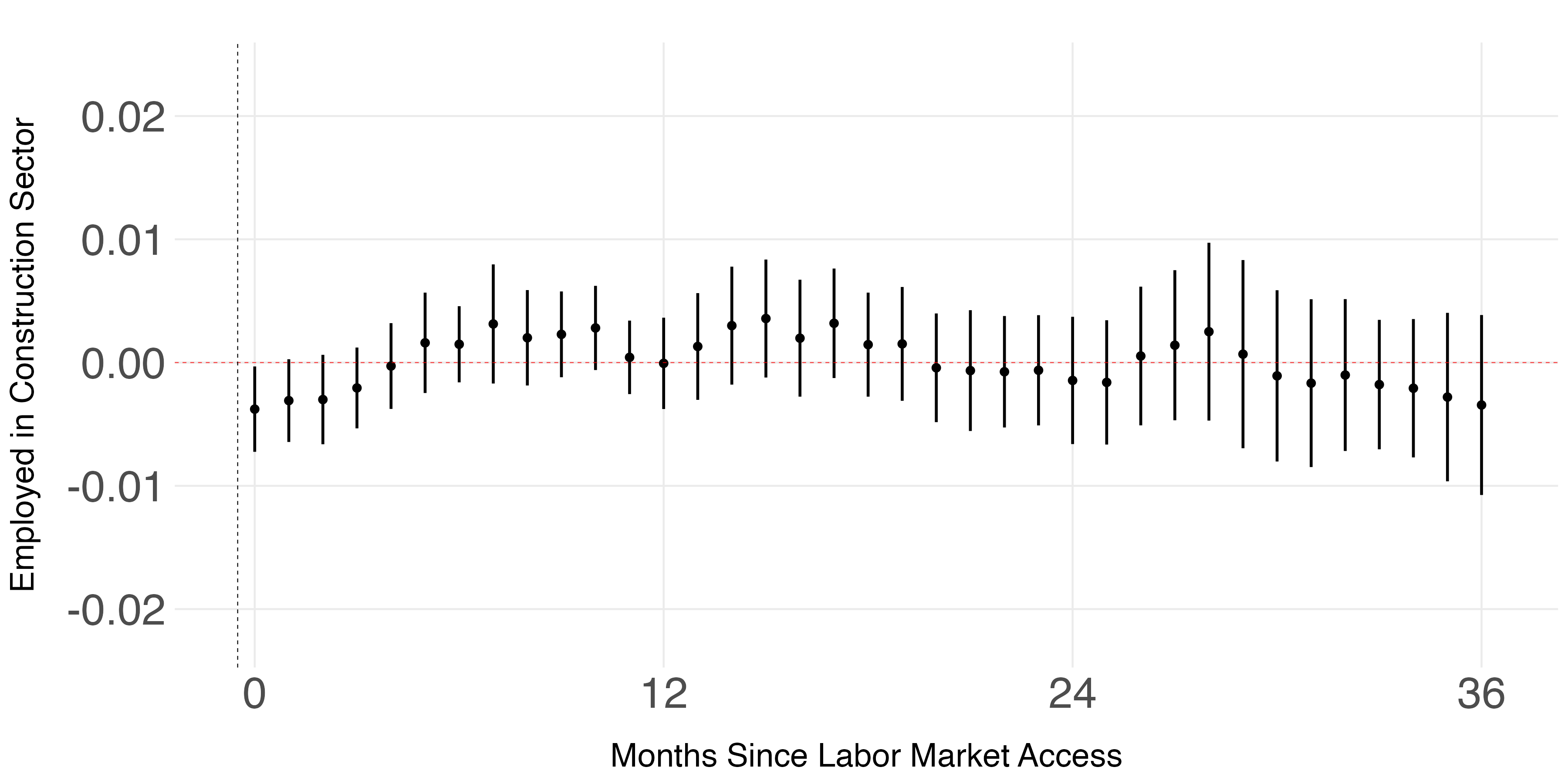}
    \label{fig:results_did_construction}
    \subcaption{Construction Sector}
\end{subfigure}
\hfill
\begin{subfigure}[b]{0.32\textwidth}
    \centering
    \includegraphics[width=\textwidth]{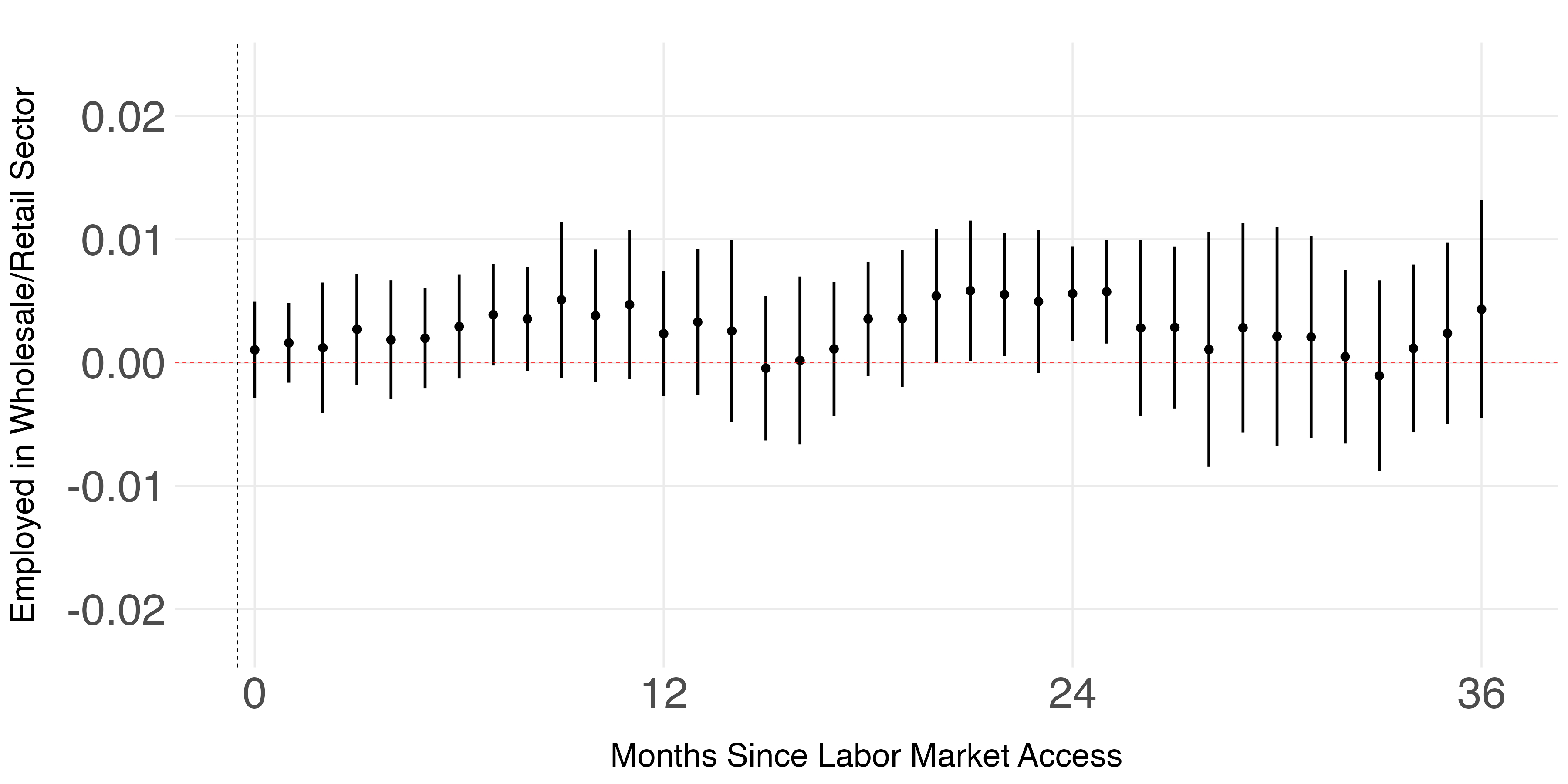}
    \label{fig:results_did_wholesale}
    \subcaption{Wholesale/Retail Sector}
\end{subfigure}

\begin{subfigure}[b]{0.32\textwidth}
    \centering
    \includegraphics[width=\textwidth]{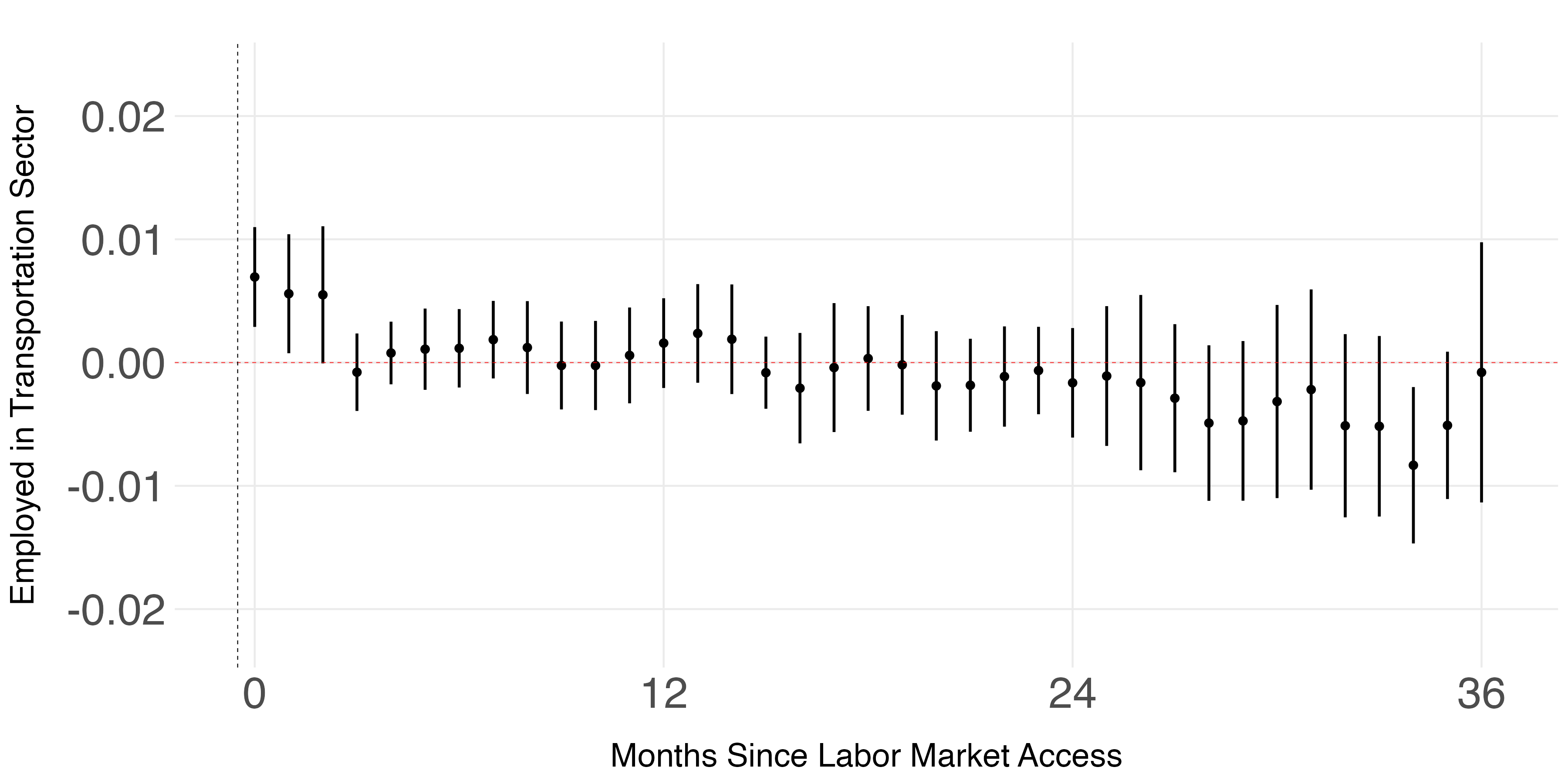}
    \label{fig:results_did_transport}
    \subcaption{Transportation Sector}
\end{subfigure}
\hfill
\begin{subfigure}[b]{0.32\textwidth}
    \centering
    \includegraphics[width=\textwidth]{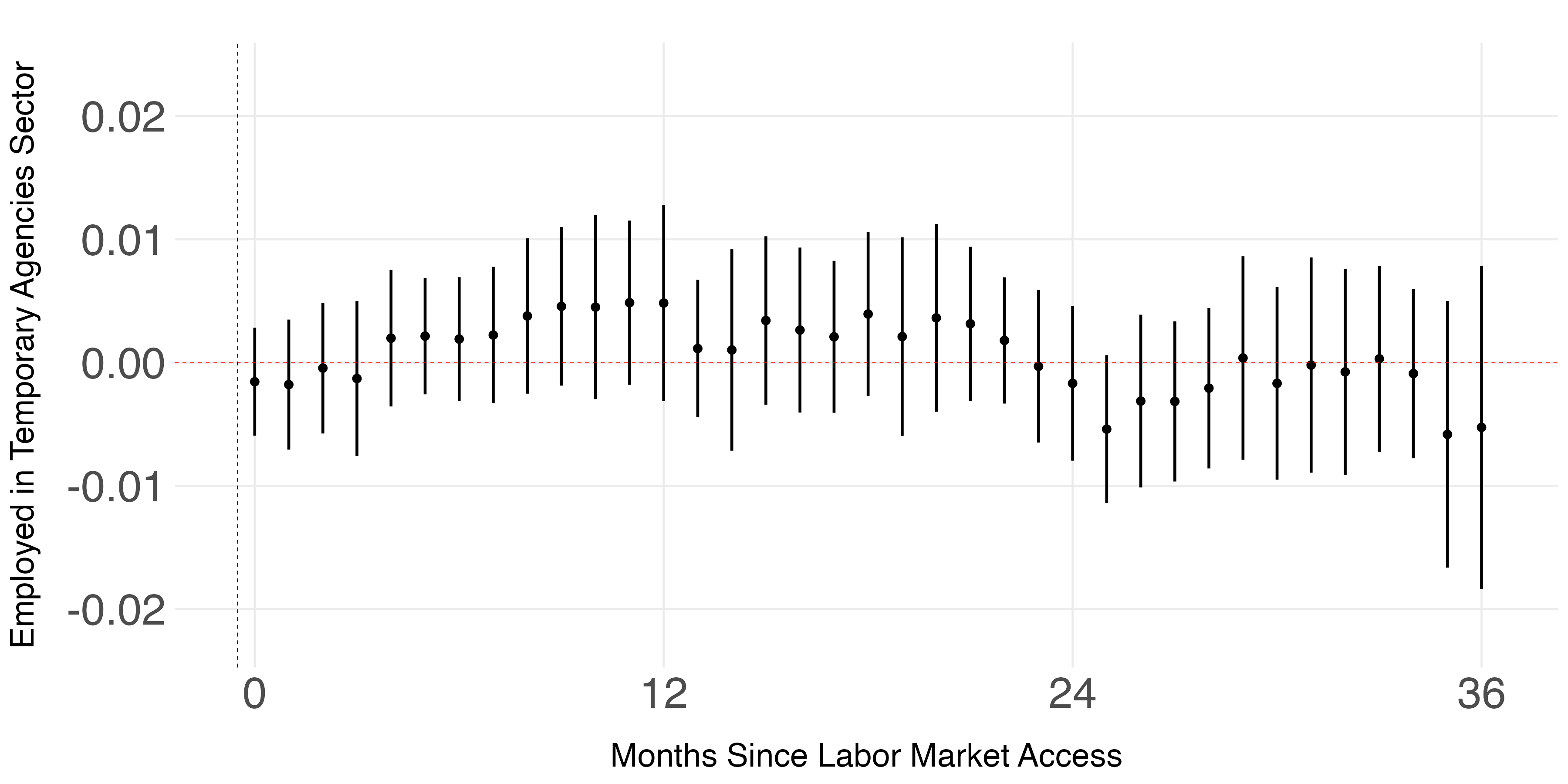}
    \label{fig:results_did_temp}
    \subcaption{Temporary Agencies Sector}
\end{subfigure}
\hfill
\begin{subfigure}[b]{0.32\textwidth}
    \centering
    \includegraphics[width=\textwidth]{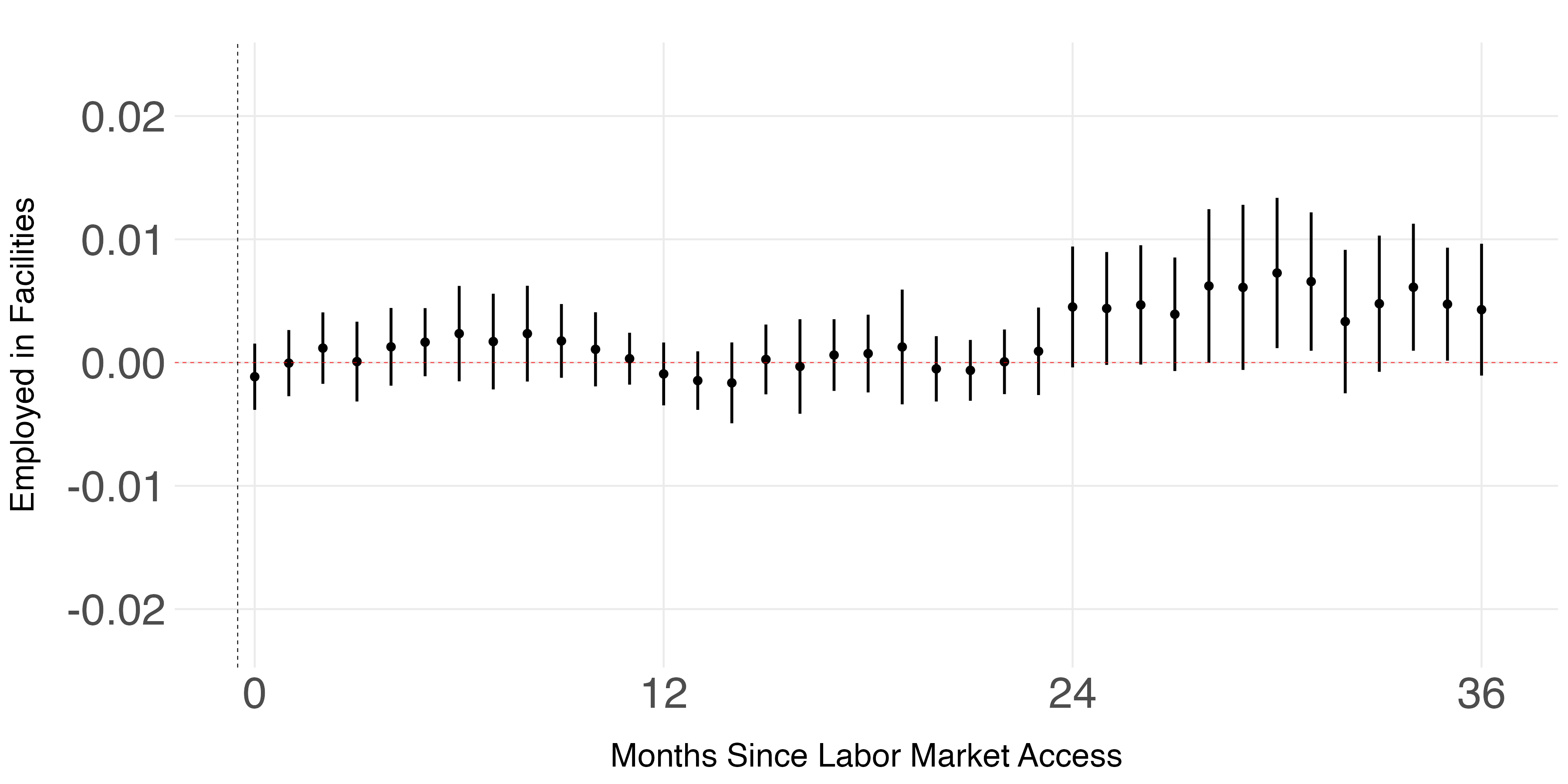}
    \label{fig:results_did_facilities}
    \subcaption{Facilities Management Sector}
\end{subfigure}

\begin{subfigure}[b]{0.32\textwidth}
    \centering
    \includegraphics[width=\textwidth]{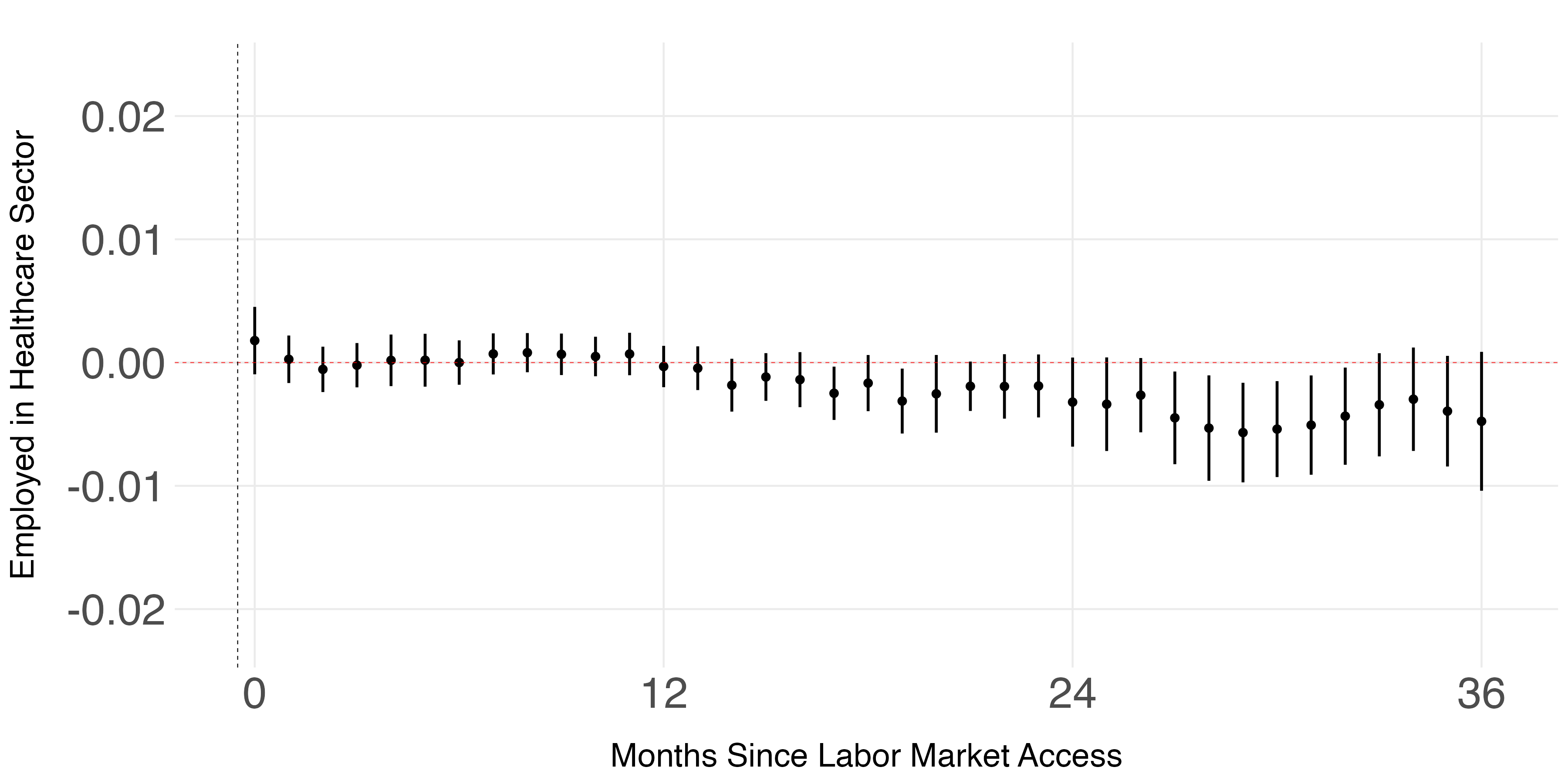}
    \label{fig:results_did_health}
    \subcaption{Healthcare Sector}
\end{subfigure}
\hfill
\begin{subfigure}[b]{0.32\textwidth}
    \centering
    \includegraphics[width=\textwidth]{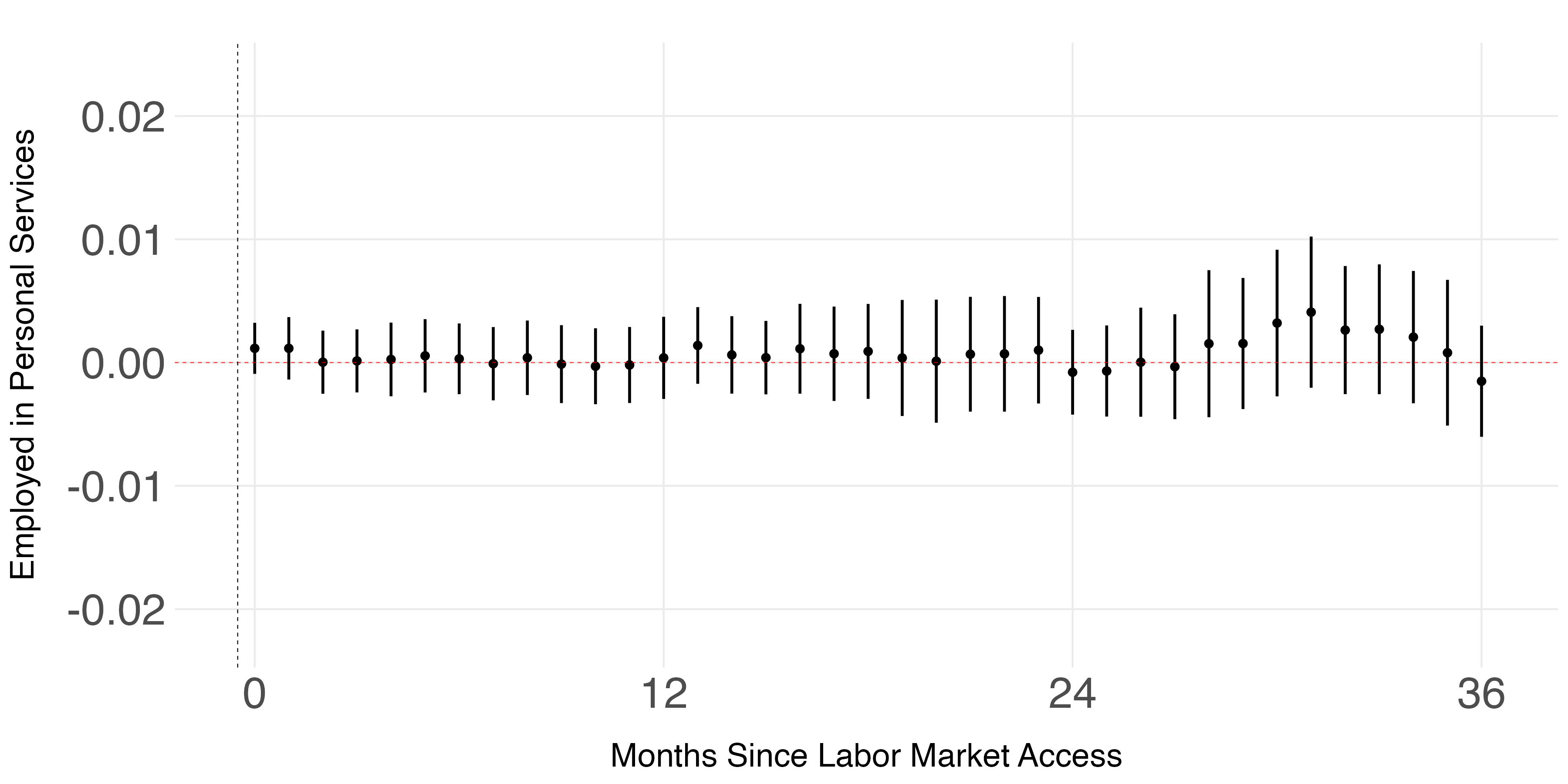}
    \label{fig:results_did_personal}
    \subcaption{Personal Service Sector}
\end{subfigure}
\hfill
\begin{subfigure}[b]{0.32\textwidth}
    \centering
    \includegraphics[width=\textwidth]{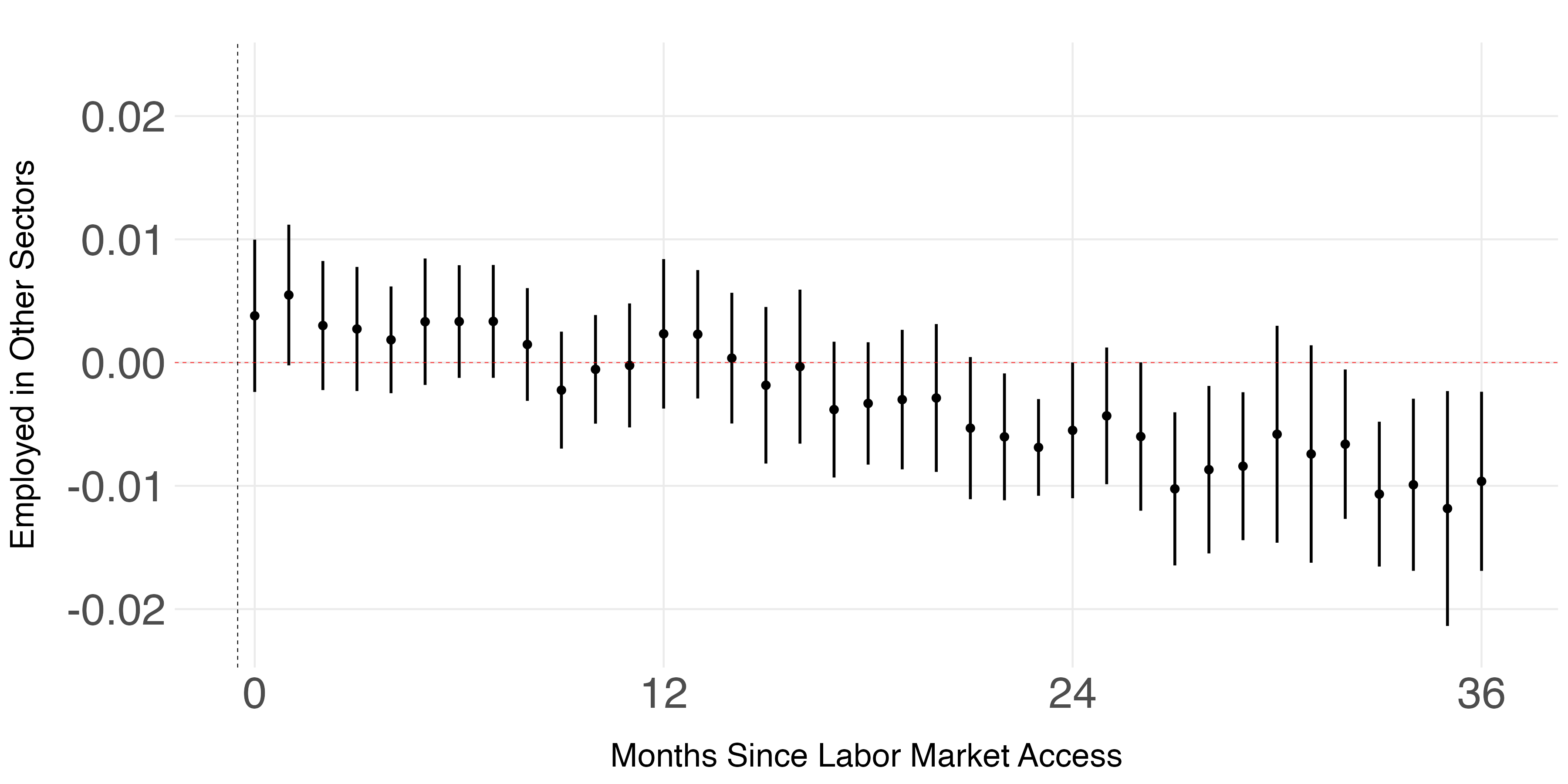}
    \label{fig:results_did_other}
    \subcaption{Other Sectors}
\end{subfigure}

\end{center}
\vspace{-3mm}
\textit{Notes}: This figure shows monthly coefficients for the $HVUR$ from estimating Equation\eqref{eq:event_study_dynamic} in my sample of refugees. Sectoral employment is defined as an individual being employed in one of the nine sectoral categories shown in Panels (a) through (i), and zero else. Sectoral categories are constructed on the following 2-digit-NACE-codes: Manufacturing (10-32), Construction (41-43), Wholesale/Retail (45-47), Transportation (49-53), Temporary Agencies (78), Facilities Management (81), Healthcare (86-88), Personal Service (96). These categories account for around 90\% of refugee employment. All other sectors are classified as "Other Sectors". Bars indicate 95\% confidence intervals.
\end{figure}

As overall employment rates converge while hospitality employment rates are elevated for treated individuals, an obvious question is which employment is crowded out for the take-up of hospitality work in the short- and medium run. To address this, I estimate equation~\eqref{eq:event_study_dynamic} with an indicator for being employed in each of the eight most relevant sectors or any other sector as the outcome variable. The most relevant sectors are those with the highest refugee employment, accounting  for around 90\% of refugee employment in the three years after labor market access. Figure~\ref{fig:results_did_sectors} shows the estimated coefficients. For most of the sectors, there are no significant differences between treatment and control group in the short-run. I interpret this as evidence that the treatment adds new employment rather than crowding out existing ones in the initial time after labor market access. The only sector that has significantly lower employment probabilities in the initial time after labor market access is the construction sector, which also has relatively low entry barriers. 

Towards the end of the first year, and possibly in response to the end of the season and  diminishing employment advantages in the hospitality sector, treated refugees are more likely to work with temporary work agencies and in facility management, two sectors that also cover a large part of overall refugee employment. Although the differences are not statistically significant, the fact that the increased take-up in these sectors coincides with the decline in seasonal job availability reveals an important mechanism. As the season ends, some treated refugees seem to transition away from the hospitality sector and into other sectors that are quite typical for refugees and that also have low entry barriers, suggesting that the refugee labor market for the most common industries is quite transitive.

Panel (i) of Figure~\ref{fig:results_did_sectors} shows the coefficients for the employment probability in sectors less common to refugees. In this category, the most important industries are the security sector, business support activities and waste collection, though no sector is particularly driving the results. In the third year, the likelihood to work in sectors of this category is significantly lower for treated refugees. This result suggests that treated individuals end up in sectors that are more common to refugees.

\paragraph{Labor market segregation} The sectoral patterns documented above suggest that early hospitality employment persistently shifts refugees toward a narrower set of industries. I quantify the aggregate implications for labor market segregation using the Duncan segregation index \citep{duncan1955methodological}, widely used to study occupational segregation by gender and space \citep{sloane2021college,campo2025political}, which compares the share of refugees in each 2-digit NACE industry to the share of natives. The sum over all industries lies between 0 (no segregation) and 1 (complete segregation). It can be interpreted as the share of refugees who would have to move sectors in order to yield no sectoral segregation. In my sample, this index is roughly 50.1\%.\footnote{In comparison to  earlier calculations and different settings, this index is higher but has a reasonable magnitude. For example, the index for migrants vs. natives with regards to occupations has been calculated as 37\% in Austria \citep{de2012immigration}} I then estimate equation~\eqref{eq:event_study_static} separately with an indicator to work in each 2-digit-NACE-industry as an outcome variable. As the resulting coefficient gives the percentage change in the likelihood to be employed in this industry, I use those coefficients to calculate the hypothetical refugee employment share for each industry in the third year after the treatment. This gives an idea of the effect on labor market segregation if all refugees would be treated. The hypothetical change from the treatment increases the Duncan segregation index to around 51.3\%. With standard errors obtained with the delta method, this is statistically significant at the 1\% level. Excluding the hospitality sector from the calculation of the index yields similar yet weaker results, as the treatment would increase the index from 43.9\% to 44.5\%. I interpret this as suggestive evidence for an increase in labor market segregation through the treatment.

Together with the lower initial employment rates of the control group, one possible explanation for this results is an increased job search effort in the control group, while the observable human capital investment through trainings do not significantly differ. Whether or not this improves refugee labor market integration ultimately depends on job quality and earnings. To capture the differences between short- and medium term career effects of the treatment, I estimate effects for the first and the third year separately. 

\subsection{The short- and medium-term effects on employment, earnings, and job characteristics}

\paragraph{First year effects} Panel A of Table~\ref{table:shortlong} shows results for the first year after labor market access. As expected, treated refugees are significantly more likely to work in the hospitality sector. Importantly, the positive effect on overall employment is even larger in magnitude and persists when excluding hospitality employment. This suggests that hospitality job availability generates spillovers into adjacent sectors: towards the end of the first year, treated refugees are more likely to work for temporary work agencies, in facility management, and in food delivery, all sectors that plausibly absorb workers transitioning out of hospitality as the season winds down.

These employment gains translate directly into higher cumulated employment (around 0.4 months after one year) and higher cumulated earnings (around \EUR{510}, or 13\% of the mean). The earnings difference is driven entirely by the employment difference rather than wage differences as average wages are not significantly different between treatment and control group. Treated refugees also do not start in lower- or higher-quality firms, as measured by AKM firm fixed effects, and show no significant differences in job stability. However, treated refugees are more likely to work in firms with lower Austrian coworker shares, an early sign of labor market segregation.

\paragraph{Third year effects} Panel B of Table~\ref{table:shortlong} shows that the higher likelihood to work in the hospitality sector persists into the third year, while overall employment rates have fully converged. Treated refugees are therefore increasingly concentrated in hospitality relative to other sectors, particularly in  healthcare and in sectors less typical for refugees. Despite converging employment rates, cumulated earnings remain significantly higher after three years (around \EUR{1,105}), again driven by cumulated employment rather than wages. Firm quality, job stability, and wages are not significantly different between the two groups, but treated refugees continue to work in firms with lower Austrian coworker shares.

Is the hospitality sector a stepping stone for refugees? The short- and medium-term results together suggest a consistent pattern: early access to hospitality work accelerates employment and raises cumulated earnings, without reducing job quality or crowding out human capital investment. The main caveat is sectoral persistence as treated refugees remain more concentrated in hospitality and refugee-typical sectors years later, and work in firms with lower Austrian coworker shares. Whether this constitutes a stepping stone therefore depends on how one weighs the earnings gains against the increased labor market segregation.

\subsection{Heterogeneity}\label{sub:heterogeneity}

Whether early hospitality work constitutes a stepping stone may also depend on who benefits. Refugees working in the hospitality sector are, on average, slightly younger and experienced longer asylum processing times (see Table~\ref{table:summary_stats_refugees}). Younger refugees may be more credit constrained and thus more responsive to immediate low-barrier work opportunities \citep{degenhardtnimczik}, but may also adapt faster and face fewer family constraints. Refugees with prolonged asylum procedures are most employment-deprived upon labor market entry, making early job availability potentially more consequential for this group \citep{hainmueller2016}. A separate question is whether the segregation effects are confined to refugees with limited labor market potential, those for whom hospitality work may be the best available option regardless, or whether they extend to refugees who could plausibly access better jobs. I examine heterogeneous effects along these dimensions.

Panel A of Table~\ref{table:heterogeneity} shows that among refugees with above-median asylum procedure, the increased hospitality take-up translates into a significantly higher overall employment likelihood in the medium term. Cumulated employment and earnings differences are also larger in magnitude than for the overall sample, without significant differences in wages, job stability, or firm quality. Early low-barrier job availability thus appears most consequential for the most employment-deprived refugees.

Panel B shows that among the group of younger refugees (aged 28 or below), effects of initial availablity of hospitality jobs on the likelihood to be employed in the hospitality sector are lasting and significant, but no significant effect on overall employment. Among younger refugees, treated individuals accumulate significantly higher earnings after three years, averaging around \EUR{1260}, without significant differences in average wages but rather better job stability and firm quality. 

While the employment and earnings gains are largest for the most employment-deprived refugees and broadly present among younger workers, it is less clear whether the sectoral persistence is similarly concentrated among those with fewer outside options, or whether it extends to refugees who could plausibly have accessed better jobs. To test this, I focus only on those with higher predicted employment probability. While education would be the natural proxy for labor market prospects, the AMDB does not contain this information. I therefore proxy employment potential by predicted employment probability after one year, constructed via an elastic net using only pre-determined characteristics: age at entry, process duration, region of origin, location of labor market access, and year of labor market access.\footnote{I use GLM with cross-validation and split the sample at the median predicted probability. The most important predictors of employment after one year are asylum process duration, initial location (in particular Vienna and Salzburg), and region of origin.}

Panel C shows that among refugees with above-median predicted employment probability, i.e. those with arguably better outside options, treated individuals are significantly more likely to work in the hospitality sector and significantly less likely to work outside it, with no significant overall employment or earnings differences. The lower Austrian coworker share persists for this group as well. Consistent with the full sample results, the sectoral decomposition also points to higher employment probability in facility management and lower probability in sectors less typical for refugees, though these results are not reported in the table.\footnote{Sectoral employment coefficients for both above- and below-median predicted employment probability groups point in a similar direction and are available upon request.} The segregation effects are thus not confined to refugees with limited labor market potential, suggesting that early hospitality work shapes labor market trajectories regardless of underlying employment prospects.

Taken together, the heterogeneity analysis adds nuance to the stepping stone question. Within the group of refugees with prolonged asylum procedures, treated individuals show the strongest overall employment gains in the medium term. Among younger workers, treated individuals show similar earnings gains compared to the full sample. Strikingly, the labor market segregation effects are also present for refugees with higher employment potential, indicating that early availability of hospitality jobs also affects the sectoral trajectories of those who could plausibly have ended up in other jobs. Whether early hospitality work constitutes a stepping stone thus depends not only on how one weighs earnings gains against segregation, but also on the group considered and the outcome examined.

\subsection{Sensitivity checks}\label{sec:robustness}

To provide evidence for the validity of my results I conduct several robustness checks.

First, I address sample imbalance. Because I restrict the analysis to pre-2020 years to exclude confounding effects from the COVID-19 pandemic, longer-term outcomes are estimated primarily from earlier cohorts (2015-2016). Panel A of Table~\ref{table:robustness} addresses this directly by restricting the sample to those receiving labor market access in 2016 or earlier, ensuring three full years of outcome observation. The results are robust to this restriction. 

Panel B addresses potential selection through geographic mobility. Although I find no significant differences in likelihood to move to Vienna, individuals might strategically relocate based on unobserved factors. Restricting the sample to non-movers generates very similar results. 

Panel C restricts the sample to individuals without pre-asylum movement. As discussed in section~\ref{sec:institutional_setting}, limited exceptions allow asylum seekers to change their location prior to receiving asylum. Excluding those who made use of these exceptions does not change the interpretation of the results. 

Panel D addresses the concern that the results may be driven by those who made use of the limited exceptions to work, e.g. in seasonal work. Results remain unchanged when excluding the few individuals utilizing these provisions, confirming that pre-treatment employment does not drive my findings.

Similarly, Panel E excludes those who took part in active labor market training or started vocational training before receiving labor market access. The latter was possible for younger individuals under certain restrictions. The interpretation of the results remains unchanged when excluding those.

Panel F excludes individuals with higher likelihood to be granted subsidiary protection rather than asylum. This addresses the concern that the timing of the subsidiary protection decision may not be captured correctly in the data. Since I cannot observe which type of protection is granted, I exclude the groups with the highest share of subsidiary protection, namely individuals from Afghanistan and Somalia \citep[see][]{steinmayr}. Again, the results do not change much.

Next, I estimate a more restrictive version of equation~\eqref{eq:event_study_static} by including location-by-year-of-entry-by-region-of-origin fixed effects. This effectively compares individuals from the same region of origin who receive labor market access within the same year and labor market region and addresses concerns about different effects for individuals from different regions. Panel G shows that, though the results become weaker, the conclusion does not change. 

I further address the concern that the treatment variable is endogenous. While the quasi-random allocation of refugees and the uncertainty in asylum procedure timing make endogeneity unlikely, one remaining concern is that regional authorities may strategically time asylum decisions to coincide with periods of high hospitality labor demand, for example to facilitate refugee employment. I already provide evidence against this in Table~\ref{tab:validating_treatment_number}, where neither the number nor the characteristics of refugees receiving labor market access correlate with the treatment variable. As a further check, in Panel H of Table~\ref{table:robustness}, I instrument the $HVUR$ with the $HVUR$ measured in the years 2010 to 2013, prior to the observation period and the large refugee inflows. This pre-period seasonality is highly correlated with current seasonality as indicated by the large first-stage F-statistic, but cannot be influenced by strategic behavior related to refugee placement. Instrumenting my treatment variable does not change the interpretation of my results.

Finally, I provide additional evidence on the role of seasonal timing by comparing outcomes for refugees receiving labor market access shortly before versus shortly after a seasonal peak by shifting the treatment variable by two months in each direction. Figure~\ref{fig:did_before_after} Panel (a) shows that refugees entering just before a peak show somewhat delayed positive effects on hospitality employment, consistent with catching the season after a short delay, while those entering just after (Panel b) show negative initial effects, consistent with missing the peak entirely. Both cases support the interpretation that the availability of hospitality jobs at the time of labor market access shapes early employment take-up.

\section{Discussion and conclusion}

This paper examines the causal impact of the early but temporary availability of low-barrier employment opportunities in the hospitality sector on refugee labor market integration in Austria. The identification strategy exploits quasi-random allocation of asylum seekers across Austrian regions combined with employment and mobility restrictions during asylum procedures. Upon receiving positive asylum decisions, refugees gain labor market access at different points within the hospitality sector's seasonal cycle, an important entry point to the labor market. This generates plausibly exogenous variation in job availability within the location and year of labor market access.

My empirical findings reveal a nuanced picture of how early low-barrier employment opportunities shape labor market integration trajectories. Seasonal hospitality jobs provide immediate employment opportunities for refugees entering Austria's labor market for the first time, with initial employment probabilities increasing substantially when refugees receive labor market access during high seasons. These initial advantages decrease as employment rates converge completely after approximately 18 months, indicating that initial job take-up alone does not generate lasting employment advantages, on average. Due to the higher cumulated employment, refugees with earlier access to hospitality jobs however generate significantly higher total earnings in the first three years, without their job quality being negatively affected.

These results complement the literature on the determinants for refugee labor market integration. Similar to \citet{degenhardtnimczik} and \citet{steinmayr}, I also find that the initial availability of low-barrier employment accelerates initial job finding. Given that the control group faces the same hospitality work opportunities merely months after or before the treatment group, this result is striking. Also in line with previous findings, this initial advantage does not result in lasting employment benefits for refugees. 

In comparison to findings by \citet{degenhardtnimczik} and \citet{arendt2022}, the treatment does however not distort human capital investments and does not affect medium-term job quality. As cumulated employment and earnings are elevated for the treatment group in the medium run, the early labor market access thus appears, at most, beneficial for refugees in net terms.

As discussed above, these findings are consistent with the notion that the integration effects of low-barrier employment depend on the job quality gap between early entry jobs and the counterfactual. In Austria's segmented refugee labor market, where low formal education, limited skill recognition, and strict occupational regulation constrain most refugees to a narrow set of low-barrier sectors, hospitality jobs may simply not constitute a fundamentally worse option than what refugees would otherwise take. This contrasts with the gig economy setting of \citet{degenhardtnimczik}, where the quality gap is potentially larger and crowding-out of human capital investment is more plausible.

At the same time, my results also suggest that refugees not only react strongly to the early availability of hospitality work, but they keep working in this sector also years later. Together with the finding that treated individuals are also working in sectors more typical to refugees and in firms with lower Austrian coworker shares, the increase in labor market segregation remains an important caveat.

Taken together, this study contributes to broader debates about optimal integration strategies by isolating the effects of early low-barrier job availability from persistent regional labor market characteristics. The findings reveal that early employment opportunities in the hospitality sector generate substantial short-term employment effects and lasting gains in cumulated employment and earnings. Policymakers should however also consider the effects on labor market segregation, particularly where work opportunities for refugees are concentrated in a small number of entry sectors. Given that hospitality represents a similarly important entry sector for refugees in Germany and Switzerland, these findings speak to integration debates well beyond the Austrian context.

\newpage\clearpage
\renewcommand{\baselinestretch}{1.12}
\setlength{\baselineskip}{9pt}
\phantomsection

\bibliographystyle{ecta}
\bibliography{References.bib}
\newpage

\clearpage 
\renewcommand{\baselinestretch}{1.5}
\setlength{\baselineskip}{20pt}

\section*{Figures and Tables}


\vspace{3cm}


\clearpage
\newpage

\makeatletter
\newcommand\primitiveinput[1]
{\input{#1} }
\makeatother

\begin{table}[h!]
\caption{Summary Statistics on Refugees}
\label{table:summary_stats_refugees}
\begin{center} 

\begin{tabular}{lcc}

\toprule
& (1) & (2) \\ 
\midrule
& \multirow{2}{*}{All refugees} & Refugees in \\

&  & Hospitality Sector \\
 \cmidrule(lr){2-2}  \cmidrule(lr){3-3}   
	\multicolumn{3}{l}{\textbf{Panel A. Demographics}}\\
\hspace{1em}Share Afghan & 0.28 & 0.34\\
\hspace{1em}Share Syrian & 0.43 & 0.39\\
\hspace{1em}Age at Entry (in years) & 29.21 & 26.75\\
\hspace{1em}Length of Asylum Procedure (in months) & 25.38 & 27.75\\
\multicolumn{3}{l}{\textbf{Panel B. Labor Market Outcomes}} \\ 
   \hspace{0.3cm}Share in Training / ALMP & 0.23 & 0.17\\
\hspace{0.3cm} Share Vocational Training & 0.05 & 0.02\\
\hspace{0.3cm} Share Employed & 0.32 & 0.49\\
\hspace{0.6cm} Share Employed in Hospitality Sector & 0.28 & 0.72\\
\hspace{0.6cm} Wage & 1 505.60 & 1 348.36\\
\hspace{0.6cm} Tenure & 7.02 & 6.54\\
\hspace{0.6cm} AKM Firm FEs & 2.03 & 2.02\\
\hspace{0.6cm} Job Take Ups & 0.47 & 0.87\\
\hspace{0.6cm} Cumulated Employment & 3.94 & 6.42\\
\hspace{0.6cm} Cumulated Earnings & 4 931.33 & 7 407.27\\
\hspace{0.6cm} Share Austrian Coworkers & 0.54 & 0.43\\
\hspace{0.3cm} Months to First Employment & 3.46 & 3.19\\
\multicolumn{3}{l}{\textbf{Panel C. Mobility Outcomes}} \\ 
   \hspace{0.3cm} Share moved to Vienna & 0.32 & 0.29\\
\hspace{0.3cm} Months to move to Vienna & 0.65 & 0.73\\
\midrule
Observations & 35 664 & 8 586\\

\bottomrule
\end{tabular}
\end{center} 
\textit{Notes}: This table shows summary statistics on all refugees in my sample (column 1) and refugees who work in the hospitality sector firms at least once in the three years after labor market access (column 2). Demographics are measured at labor market access. Labor market outcomes are measured one year after labor market access. The share of employed in hospitality sector, wages, tenure, AKM Firm FEs and share Austrian coworkers are conditional on being employed. Job Take ups are conditional on being employed at least once in the first year. Wages are measured in EUR per month, tenure is measured in months. The share of movers to Vienna is conditional on not receiving labor market access in Vienna.
\end{table}

\newpage

\newpage
	
\begin{table}[h!] \caption{The average effect of seasonal hospitality job availability on employment and mobility}\label{table:main_results}
\begin{center} 
\resizebox{\linewidth}{!}{%
\begin{tabular}{l*{7}{c}}
\toprule
& (1) & (2) & (3) & (4) & (5) & (6)   \\
\midrule

& \multicolumn{4}{c}{Employment} &   \multicolumn{2}{c}{Mobility} \\
 \cmidrule(lr){2-5}     \cmidrule(lr){6-7}      
& Hospitality  &   & Non-Hospitality & Months to & Move to & Months to \\
&  Firm &   Overall & Firm & First Job   & Vienna & Move  \\
 \cmidrule(lr){2-2}  \cmidrule(lr){3-3}   \cmidrule(lr){4-4}  \cmidrule(lr){5-5}  \cmidrule(lr){6-6}  \cmidrule(lr){7-7} 
   HVUR & 0.011$^{***}$ & 0.014$^{***}$ & 0.003 & -0.329$^{***}$ & 0.000 & 0.512\\
 & (0.004) & (0.005) & (0.005) & (0.060) & (0.003) & (0.366)\\
Mean of Independent Variable & 0.09 & 0.33 & 0.24 & 11.69 & 0.34 & 2.34\\
\midrule
Observations & 1,013,867 & 1,013,867 & 1,013,867 & 23,405 & 759,224 & 9,156\\

\bottomrule
\end{tabular}
}
\end{center} 
\textit{Notes}: This table shows coefficients from estimating the effect of temporary hospitality job availability on refugee labor market outcomes. Both Panels estimate equation \eqref{eq:event_study_static}. I report the coefficient for the treatment variable, $HVUR$. The outcome variables are a dummy for employment in the hospitality sector, a dummy for being employed, and a dummy for being employed outside the hospitality sector. For the latter,  employment in the hospitality sector is subtracted from employment. Further, I examine the average month to ones first job, conditional that an individual has at some point worked. Also, I show estimates an indicator for moving to Vienna (conditional on not receiving labor market access in Vienna) and the months until a move to Vienna (conditional on having moved to Vienna). All regressions control for AMS-region times year of labor market entry fixed effects, the age at the time of labor market access, the time since arrival in Austria, dummies for 21 regions of origin and the vacancy-to-unemployment ratio for non-hospitality jobs. Standard errors are clustered at the interacted labor market region (AMS-region) and year of labor market entry as well as calendar month level. Stars indicate significance levels:  *~$p<0.10$, **~$p<0.05$, ***~$p<0.01$.
\end{table}


\begin{landscape}
	
\begin{table}[h!] \caption{Short-vs. medium-term effects}\label{table:shortlong}
\begin{center} 
\resizebox{\linewidth}{!}{%
\begin{tabular}{l*{10}{c}}
\toprule
& (1) & (2) & (3) & (4) & (5) & (6) & (7) & (8) & (9) & (10) \\
\midrule
& \multicolumn{4}{c}{Employment} &   \multicolumn{2}{c}{Labor Income} & \multicolumn{4}{c}{Job Stability and Quality} \\
 \cmidrule(lr){2-5}     \cmidrule(lr){6-7}   \cmidrule(lr){8-11}        
& Hospitality  &   & Non-Hospitality & Cumulated &  log & Cumulated &  & New & Coworker Share  &  AKM \\
&  Firm &   Overall & Firm       & Employment   & Wage & Earnings & Tenure & Job & Austrians & Firm FE \\
 \cmidrule(lr){2-2}  \cmidrule(lr){3-3}   \cmidrule(lr){4-4}  \cmidrule(lr){5-5}  \cmidrule(lr){6-6}  \cmidrule(lr){7-7}  \cmidrule(lr){8-8}  \cmidrule(lr){9-9} \cmidrule(lr){10-10} \cmidrule(lr){11-11} 
\multicolumn{10}{l}{\textbf{Panel A: Average effects in the first year}} \\
\midrule
HVUR & 0.010$^{**}$ & 0.021$^{***}$ & 0.011$^{**}$ & 0.406$^{***}$ & 0.005 & 510.893$^{**}$ & 0.019 & 0.002 & -0.009$^{**}$ & 0.002\\
 & (0.004) & (0.005) & (0.004) & (0.150) & (0.007) & (200.505) & (0.056) & (0.001) & (0.003) & (0.003)\\
\midrule
Observations & 402,344 & 402,344 & 402,344 & 36,061 & 101,445 & 36,061 & 101,445 & 402,344 & 88,802 & 89,253\\
 
\midrule \\
\multicolumn{10}{l}{\textbf{Panel B: Average effects in the third year}}\\
HVUR & 0.015$^{***}$ & 0.007 & -0.007 & 0.639$^{**}$ & 0.010 & 1 105.530$^{**}$ & 0.125 & 0.000 & -0.012$^{***}$ & 0.002\\
 & (0.004) & (0.008) & (0.008) & (0.293) & (0.007) & (440.377) & (0.155) & (0.001) & (0.004) & (0.003)\\
\midrule
Observations & 272,143 & 272,143 & 272,143 & 25,203 & 116,615 & 25,203 & 116,615 & 272,143 & 99,060 & 98,191\\

\bottomrule
\end{tabular}
}
\end{center} 
\textit{Notes}: This table shows coefficients from estimating the effect of temporary hospitality job availability on refugee labor market outcomes in the first and third year, respectively. Both Panels estimate equation~\eqref{eq:event_study_static}. I report the coefficient for the treatment variable, $HVUR$ for each subsample. The outcome variables are a dummy for employment in a hospitality firm, a dummy for being employed, and a dummy for being employed outside the hospitality sector. For the latter,  employment in the hospitality sector is subtracted from employment. Further, I examine cumulated employment and cumulated earnings, measured in the last month of each sample, monthly log wages (conditional on working), tenure in months (conditional on working), an indicator for starting a new job, the share of coworkers who are Austrian (conditional on working), AKM firm FEs (conditional on working). Both the coworker share of Austrians and AKM firm FEs are based on the calculation in \citet{degenhardtnimczik}, and missing values arise due to sample restrictions, e.g. with respect to firm size. The control variables correspond to those in Table~\ref{table:main_results}. Standard errors are clustered at the interacted labor market region (AMS-region) and year of labor market entry as well as calendar month level. Stars indicate significance levels:  *~$p<0.10$, **~$p<0.05$, ***~$p<0.01$.
\end{table} 

\end{landscape}

\begin{landscape}
	
\begin{table}[h!] \caption{Heterogeneity for medium-term effects}\label{table:heterogeneity}
\begin{center} 
\resizebox{0.77\linewidth}{!}{%
\begin{tabular}{l*{10}{c}}
\toprule
& (1) & (2) & (3) & (4) & (5) & (6) & (7) & (8) & (9) & (10) \\
\midrule
& \multicolumn{4}{c}{Employment} &   \multicolumn{2}{c}{Labor Income} & \multicolumn{4}{c}{Job Stability and Quality} \\
 \cmidrule(lr){2-5}     \cmidrule(lr){6-7}   \cmidrule(lr){8-11}      
& Hospitality  &   & Non-Hospitality & Cumulated &  log & Cumulated &  & New & Coworker Share  &  AKM \\
&  Firm &   Overall & Firm       & Employment   & Wage & Earnings & Tenure & Job & Austrians & Firm FE \\
 \cmidrule(lr){2-2}  \cmidrule(lr){3-3}   \cmidrule(lr){4-4}  \cmidrule(lr){5-5}  \cmidrule(lr){6-6}  \cmidrule(lr){7-7}  \cmidrule(lr){8-8}  \cmidrule(lr){9-9} \cmidrule(lr){10-10} \cmidrule(lr){11-11} 

\multicolumn{10}{l}{\textbf{Panel A: Long ($>$ median) asylum process duration}}\\
\midrule
HVUR & 0.028$^{***}$ & 0.021$^{**}$ & -0.007 & 0.982$^{***}$ & -0.001 & 1 333.931$^{**}$ & 0.095 & 0.001 & -0.012 & 0.003\\
 & (0.007) & (0.009) & (0.008) & (0.306) & (0.010) & (571.392) & (0.210) & (0.002) & (0.008) & (0.005)\\
\midrule
Observations & 129,937 & 129,937 & 129,937 & 12,377 & 63,615 & 12,377 & 63,615 & 129,937 & 55,437 & 54,985\\

\midrule \\
\multicolumn{10}{l}{\textbf{Panel B: Young ($\leq$ 28) }} \\
\midrule
HVUR & 0.017$^{*}$ & 0.010 & -0.007 & 0.720$^{**}$ & 0.016 & 1 257.708$^{**}$ & 0.384 & -0.004$^{**}$ & -0.013$^{*}$ & 0.009\\
 & (0.009) & (0.012) & (0.013) & (0.328) & (0.014) & (471.824) & (0.282) & (0.002) & (0.008) & (0.005)\\
\midrule
Observations & 131,913 & 131,913 & 131,913 & 12,336 & 61,558 & 12,336 & 61,558 & 131,913 & 56,969 & 56,533\\
 
\midrule \\
\multicolumn{10}{l}{\textbf{Panel C: High predicted employment propensity}}\\
\midrule
HVUR & 0.031$^{***}$ & 0.007 & -0.024$^{**}$ & 0.553$^{*}$ & 0.001 & 645.426 & -0.062 & -0.001 & -0.016$^{**}$ & -0.001\\
 & (0.010) & (0.011) & (0.011) & (0.310) & (0.014) & (590.657) & (0.278) & (0.002) & (0.008) & (0.006)\\
\midrule
Observations & 89,393 & 89,393 & 89,393 & 9,160 & 51,835 & 9,160 & 51,835 & 89,393 & 47,484 & 46,802\\

\bottomrule
\end{tabular}
}
\end{center} 
\textit{Notes}: This table shows coefficients from estimating the effect of temporary hospitality job availability on refugee labor market outcomes in different subgroups of the male refugee population. I estimate equation\eqref{eq:event_study_static} and report the coefficient for the treatment variable, $HVUR$. Panel A shows the coefficients for refugees aged 28 or younger at the time of labor market access. 28 is the median age at labor market access. Panel B shows coefficients for refugees with process duration above the median length of the asylum procedure. The median length is calculated separately for each labor market region (AMS-region) and year of entry. Panel C shows coefficients for the group of refugees with above median predicted employment probability as described in section~\ref{sub:heterogeneity}. Differences in observation counts across panels reflect differences in the distribution of entry cohorts across subgroups, which affects the number of individuals observed for the full three-year period given the pre-2020 sample restriction. The outcome and control variables correspond to those in Table~\ref{table:shortlong}.  Standard errors are clustered at the interacted labor market region (AMS-region) and year of labor market entry as well as calendar month level. Stars indicate significance levels:  *~$p<0.10$, **~$p<0.05$, ***~$p<0.01$.

\end{table} 

\end{landscape}

\begin{table}[h!] \caption{Sensitivity}\label{table:robustness}
\begin{center} 
\resizebox{0.73\linewidth}{!}{%
\begin{tabular}{l*{7}{c}}
\toprule
& (1) & (2) & (3) & (4) & (5) & (6)   \\
\midrule

& \multicolumn{4}{c}{Employment} &   \multicolumn{2}{c}{Mobility} \\
 \cmidrule(lr){2-5}     \cmidrule(lr){6-7}      
& Hospitality  &   & Non-Hospitality & Months to & Move to & Months to \\
&  Firm &   Overall & Firm & First Job   & Vienna & Move  \\
 \cmidrule(lr){2-2}  \cmidrule(lr){3-3}   \cmidrule(lr){4-4}  \cmidrule(lr){5-5}  \cmidrule(lr){6-6}  \cmidrule(lr){7-7}

\multicolumn{7}{l}{\textbf{Panel A: Labor Market access $\leq$ 2016}} \\
HVUR & 0.012$^{**}$ & 0.013$^{**}$ & 0.001 & -0.600$^{***}$ & 0.001 & 0.731\\
 & (0.005) & (0.006) & (0.007) & (0.185) & (0.005) & (0.578)\\
\midrule
Observations & 643,249 & 643,249 & 643,249 & 11,899 & 506,860 & 6,576\\

\midrule \\

\multicolumn{7}{l}{\textbf{Panel B: Only stayers}}\\

HVUR & 0.012$^{***}$ & 0.014$^{**}$ & 0.002 & -0.481$^{***}$ &  & \\
 & (0.004) & (0.006) & (0.006) & (0.070) &  & \\
\midrule
Observations & 640,232 & 640,232 & 640,232 & 17,026 &  & \\
 
\midrule \\

\multicolumn{7}{l}{\textbf{Panel C: No pre-movers}}\\

HVUR & 0.010$^{***}$ & 0.013$^{**}$ & 0.003 & -0.309$^{***}$ & 0.000 & 0.557\\
 & (0.004) & (0.005) & (0.005) & (0.053) & (0.004) & (0.341)\\
\midrule
Observations & 912,080 & 912,080 & 912,080 & 21,355 & 730,772 & 8,837\\
 
\midrule \\

\multicolumn{7}{l}{\textbf{Panel D: No pre-employed}}\\

HVUR & 0.011$^{***}$ & 0.015$^{***}$ & 0.004 & -0.379$^{***}$ & 0.000 & 0.542\\
 & (0.004) & (0.005) & (0.005) & (0.065) & (0.004) & (0.395)\\
\midrule
Observations & 934,846 & 934,846 & 934,846 & 20,818 & 703,659 & 8,484\\
 
\midrule \\

\multicolumn{7}{l}{\textbf{Panel E: No pre-training}}\\

HVUR & 0.011$^{**}$ & 0.014$^{**}$ & 0.004 & -0.411$^{***}$ & 0.002 & 1.059$^{*}$\\
 & (0.005) & (0.006) & (0.006) & (0.065) & (0.003) & (0.563)\\
\midrule
Observations & 574,503 & 574,503 & 574,503 & 13,141 & 432,070 & 3,387\\
 
\midrule \\

\multicolumn{7}{l}{\textbf{Panel F: Only those with low likelihood of subsidiary protection}}\\

HVUR & 0.011$^{**}$ & 0.014$^{**}$ & 0.003 & -0.344$^{***}$ & -0.004 & 0.554\\
 & (0.004) & (0.006) & (0.006) & (0.042) & (0.003) & (0.457)\\
\midrule
Observations & 744,716 & 744,716 & 744,716 & 15,557 & 549,349 & 6,660\\
 
\midrule \\

\multicolumn{7}{l}{\textbf{Panel G: Additional region-of-origin FE interaction}} \\
HVUR & 0.009$^{**}$ & 0.011$^{**}$ & 0.003 & -0.303$^{***}$ & 0.000 & 0.365\\
 & (0.004) & (0.005) & (0.005) & (0.059) & (0.004) & (0.416)\\
\midrule
Observations & 1,013,867 & 1,013,867 & 1,013,867 & 23,405 & 759,224 & 9,156\\

\midrule \\

\multicolumn{7}{l}{\textbf{Panel H: Instrumented treatment variable}}\\

HVUR & 0.009$^{**}$ & 0.014$^{**}$ & 0.004 & -0.337$^{***}$ & 0.006 & 0.546\\
 & (0.004) & (0.006) & (0.007) & (0.081) & (0.004) & (0.569)\\
First-stage F-stat & 3,442,104 & 3,442,104 & 3,442,104 & 73,486 & 2,614,585 & 21,494\\
\midrule
Observations & 1,013,867 & 1,013,867 & 1,013,867 & 23,405 & 759,224 & 9,156\\

\bottomrule
\end{tabular}
}
\end{center} 
\textit{Notes}: This table provides sensitivity checks to my main results from estimating equation~\eqref{eq:event_study_static}, shown in Table~\ref{table:main_results}. Panel A restricts the sample to those who received asylum in 2016 or earlier. Panel B restricts the sample to refugees who stay in their originally assigned state. Panel C excludes all refugees who use one of the exceptions to move their assigned location before labor market access. Panel D excludes all refugees who use one of the exceptions to be employed or self-employed before labor market access. Panel E excludes all refugees who use one of the exceptions to take up vocational training or active labor market training before labor market access. Panel F excludes all individuals from Afghanistan and Somalia, the two countries with the highest share of subsidiary protection. Panel G includes additionally interacts the year-of-entry and AMS-region fixed effects with region-of-origin fixed effects. Panel H instruments the $HVUR$ by the pre-period $HVUR$. All outcome variables and control variables correspond to those in Table~\ref{table:main_results}. Standard errors are clustered at the interacted labor market region (AMS-region) and year of labor market entry as well as calendar month level. Stars indicate significance levels:  *~$p<0.10$, **~$p<0.05$, ***~$p<0.01$.

\end{table}

\appendix
\renewcommand\thesection{\Alph{section}}
\setcounter{table}{0} \renewcommand{\thetable}{A.\arabic{table}}
\setcounter{figure}{0} \renewcommand{\thefigure}{A.\arabic{figure}}

\newpage
\clearpage 
\noindent\textbf{\Large ONLINE APPENDIX}


\section{Additional Figures and Tables}

\begin{figure}[h!]
\begin{center}
\caption{Seasonal variation refugee labor market access}
\label{fig:seasonality_refugees_vienna}

        \centering
\includegraphics[width=\textwidth]{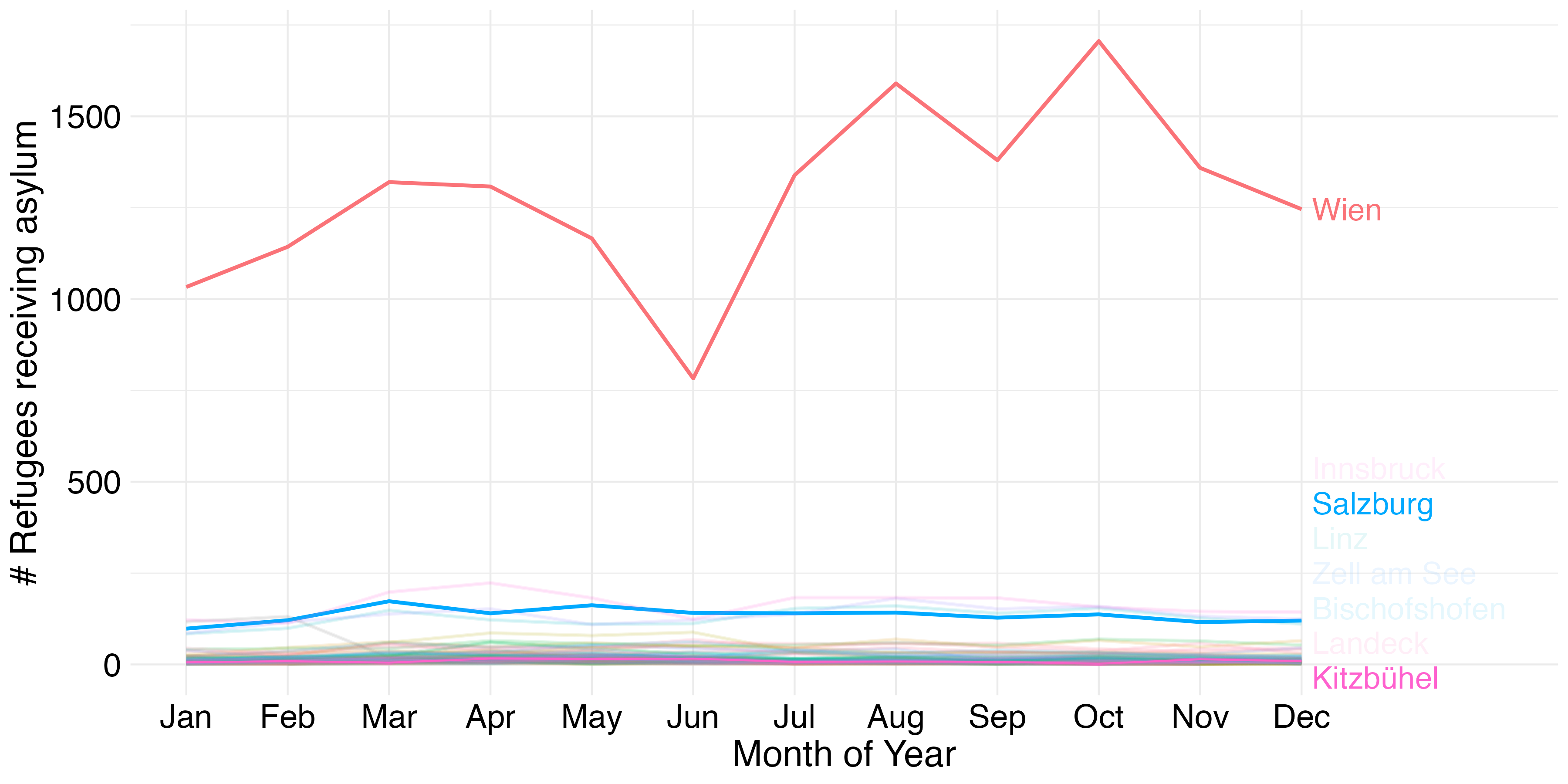}
\subcaption{Within-year variation of the number of refugees who received labor market access} 
\end{center}
 \textit{Notes}: This figure shows the seasonality of refugees' labor market access in Austria. It shows the number of refugees who received asylum. Values are aggregated and averaged on the month-of-year and AMS-region. AMS-regions for Vienna are aggregated to one region.

\end{figure}


\begin{figure}[h!]
\begin{center}
\caption{The effect of entering shortly before or shortly after high seasonal hospitality job availability on refugee employment in the hospitality sector}
\label{fig:did_before_after}

\begin{subfigure}[b]{0.8\textwidth}
        \centering
\includegraphics[width=\textwidth]{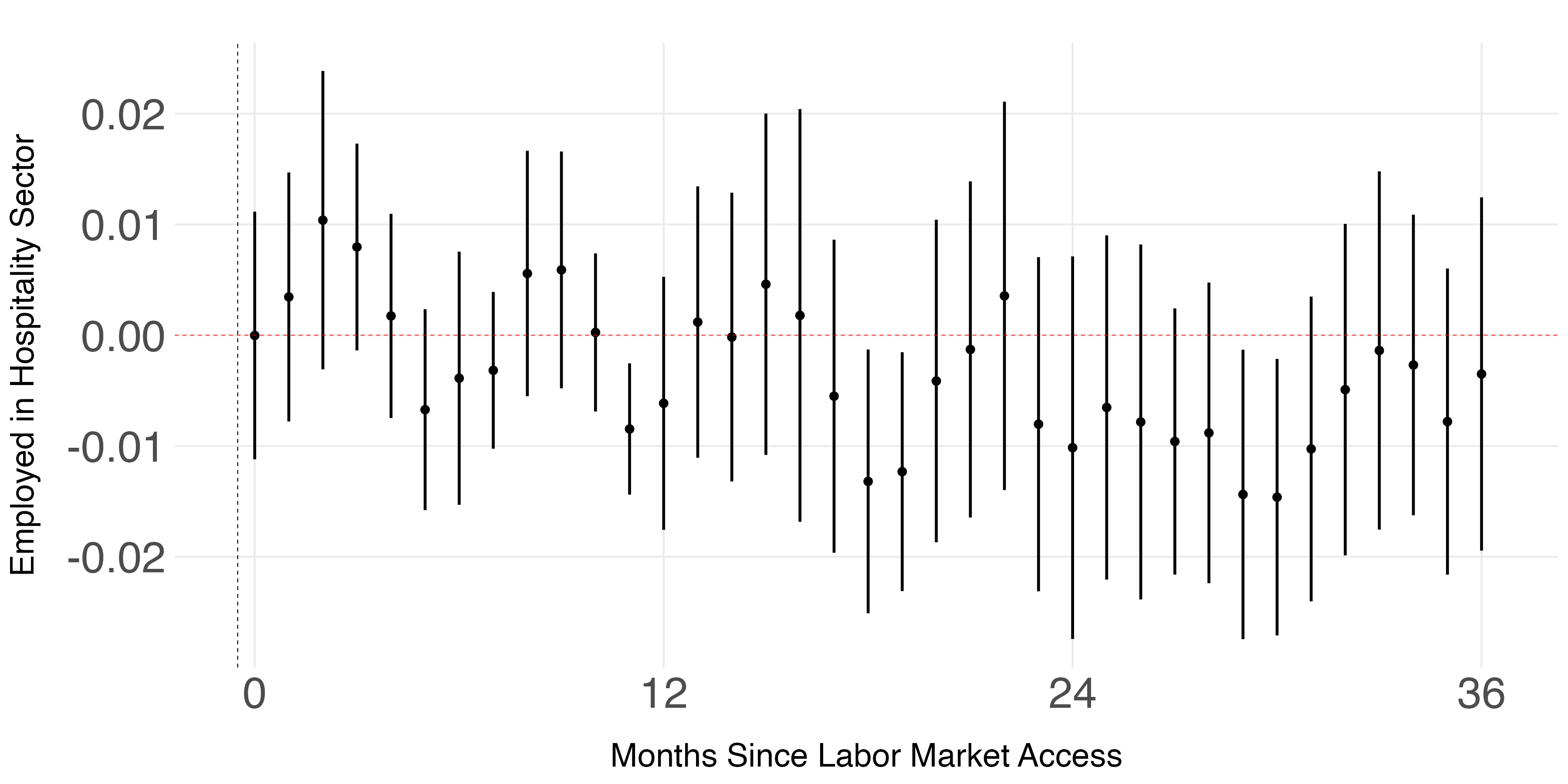}
\subcaption{Entering two months earlier}  \label{fig:did_before}
\end{subfigure}
\begin{subfigure}[b]{0.8\textwidth}
        \centering
\includegraphics[width=\textwidth]{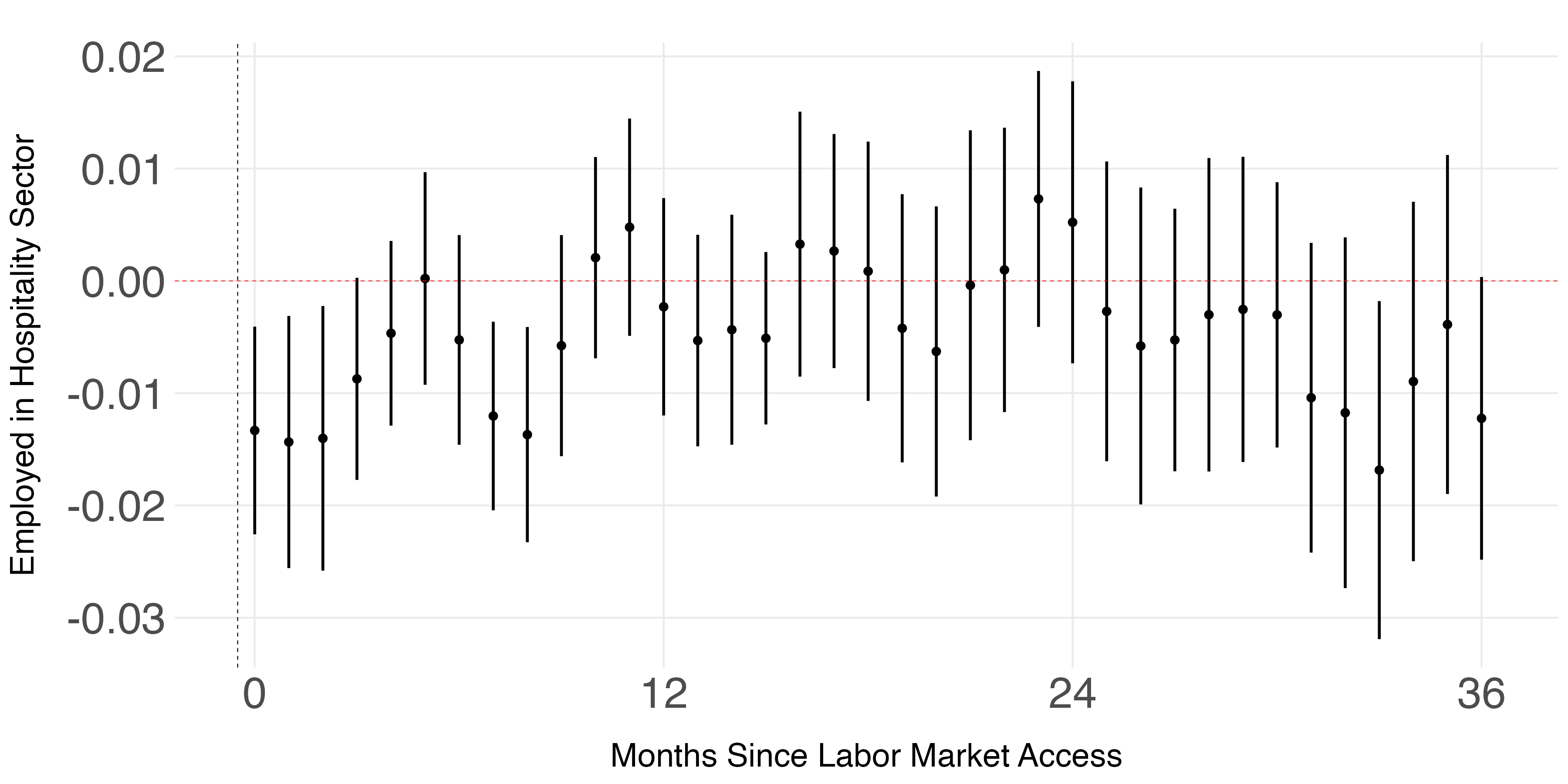} 
\subcaption{Entering two months later}  
\label{fig:did_after}
\end{subfigure}

\end{center}
  \textit{Notes}: This figure shows monthly coefficients for the $HVUR$ from estimating Equation~\eqref{eq:event_study_dynamic} in my sample of refugees, shifting the treatment variable by two months in each direction. Panel (a) plots the effect on employment in the hospitality sector for refugees receiving labor market access two months earlier than the observed $HVUR$. Panel (b) shows the corresponding effect for refugees receiving labor market access two months later than the observed $HVUR$. Bars indicate 95\% confidence intervals.
\end{figure}
\vspace{3cm}


\begin{table}[hp]
\begin{center}
 \caption{Correlation of treatment variable, number of refugees, and refugee characteristics} 
\label{tab:validating_treatment_number}
\resizebox{\linewidth}{!}{%
\centering
\begin{tabular}{lcccccccccc}
   \tabularnewline  \midrule
                         & (1) & (2) & (3) & (4) & (5) & (6) & (7) & (8) & (9) & (10)  \\  
   \midrule
                & log(Number & Number & & Average  &  & & Share & Share & Share & Share  \\  
                & of & of & Average & Process & Share & Share & pre & pre & pre & pre \\
                & Refugees) & Refugees & Age & Duration & Afghan & Syrian & Empl. & Empl. Hosp. & Self-Empl. & Schulung \\
   \midrule
   log(HVUR)    & 0.01   &        &       &        &        &      &         &        &        &         \\
                & (0.03) &        &       &        &        &      &         &        &        &         \\
   HVUR         &        & 4.28   & 1.02  & -6.55  & -0.16  & 0.06 & -0.14  & -0.02  & 0.01   & -0.13   \\
                &        & (2.90) & (4.09)& (7.05) & (0.24) & (0.19) & (0.13) & (0.12) & (0.02) & (0.09) \\
  
   \midrule
   Observations & 4,251  & 4,251  & 4,251 & 4,251  & 4,251  & 4,251 & 4,251  & 4,251  & 4,251  & 4,251  \\  
   \bottomrule
\end{tabular}
}
\end{center}
\textit{Notes}: This table examines the correlation of the treatment variable, $HVUR$, and the number as well as average characteristics of positive asylum decisions in Austrian labor market regions (AMS regions) at the time of labor market access. AMS-regions for Vienna are aggregated to one region. Column (1) estimates the elasticity of the number of refugees with respect to the HVUR using a log-log specification. Columns (2)--(10) use the level of HVUR. I regress the number of refugees and average characteristics of those male refugees who receive asylum or subsidiary protection on the treatment variable, HVUR. Observations are aggregated on the AMS-region and month of labor market access level, columns (3)--(10) are based on mean values for refugees receiving labor market access in each cell. All regressions are pooled estimations for the years from 2014 to 2019 and include interacted year and AMS-region FEs as well as calendar month FEs. Standard errors are clustered at the interacted labor market region (AMS-region) and year of labor market entry as well as calendar month level. Stars indicate significance levels:  *~$p<0.10$, **~$p<0.05$, ***~$p<0.01$. 
\end{table}

\newpage

\begin{table}[h!] \caption{The average effect of seasonal hospitality job availability on additional outcomes}\label{table:appendix_results}
\begin{center} 
\begin{tabular}{l*{4}{c}}
\toprule
& (1) & (2) & (3) & (4)   \\
\midrule
& \multicolumn{2}{c}{Training} &   \multicolumn{2}{c}{Moved to Vienna during} \\
 \cmidrule(lr){2-3}     \cmidrule(lr){4-5}      
& Labor Market & Vocational & High & Low \\
& Training & Training & Season & Season  \\
 \cmidrule(lr){2-2}  \cmidrule(lr){3-3}   \cmidrule(lr){4-4}  \cmidrule(lr){5-5} 
   HVUR & -0.005 & 0.000 & -0.002 & 0.008$^{**}$\\
 & (0.005) & (0.002) & (0.004) & (0.004)\\
\midrule
Observations & 476,624 & 1,013,867 & 89,757 & 86,768\\

\bottomrule
\end{tabular}
\end{center}
\textit{Notes}:  This table shows additional outcomes from estimating Equation~\eqref{eq:event_study_static}. The outcome variables are a dummy for labor market training participation (conditional on unemployment), a dummy for vocational training participation, and a dummy for moving to Vienna (conditional on not receiving labor market access in Vienna). Columns (3) and (4) are restricted to include the first six months after labor market access (where almost 90\% of moves take place). Column (3) restricts to months of the year where the $HVUR$ is above median in each AMS-region, column (4) below median. All control variables correspond to those in Table~\ref{table:main_results}. Standard errors are clustered at the interacted labor market region (AMS-region) and year of labor market entry as well as calendar month level. Stars indicate significance levels:  *~$p<0.10$, **~$p<0.05$, ***~$p<0.01$. 
\end{table}


\begin{landscape}
  
\begin{table}[h!] \caption{Medium-term effects for movers}\label{table:movers}
\begin{center} 
\resizebox{0.77\linewidth}{!}{%
\begin{tabular}{l*{10}{c}}
\toprule
& (1) & (2) & (3) & (4) & (5) & (6) & (7) & (8) & (9) & (10) \\
\midrule
& \multicolumn{4}{c}{Employment} &   \multicolumn{2}{c}{Labor Income} & \multicolumn{4}{c}{Job Stability and Quality} \\
 \cmidrule(lr){2-5}     \cmidrule(lr){6-7}   \cmidrule(lr){8-11}      
& Hospitality  &   & Non-Hospitality & Cumulated &  log & Cumulated &  & New & Coworker Share  &  AKM \\
&  Firm &   Overall & Firm       & Employment   & Wage & Earnings & Tenure & Job & Austrians & Firm FE \\
 \cmidrule(lr){2-2}  \cmidrule(lr){3-3}   \cmidrule(lr){4-4}  \cmidrule(lr){5-5}  \cmidrule(lr){6-6}  \cmidrule(lr){7-7}  \cmidrule(lr){8-8}  \cmidrule(lr){9-9} \cmidrule(lr){10-10} \cmidrule(lr){11-11} 
\multicolumn{10}{l}{\textbf{Only movers}}\\
\midrule
HVUR & 0.010 & 0.006 & -0.004 & 0.373 & 0.011 & 572.589 & 0.079 & -0.001 & -0.008 & -0.006\\
 & (0.009) & (0.012) & (0.012) & (0.255) & (0.015) & (493.123) & (0.197) & (0.002) & (0.008) & (0.008)\\
\midrule
Observations & 117,894 & 117,894 & 117,894 & 10,641 & 45,262 & 10,641 & 45,262 & 117,894 & 36,628 & 36,484\\

\bottomrule
\end{tabular}
}
\end{center} 
\textit{Notes}: This table shows coefficients from estimating the effect of temporary hospitality job availability on refugee labor market outcomes for the sample of refugees who eventually change their originally assigned state within the observation period. I estimate equation\eqref{eq:event_study_static} and report the coefficient for the treatment variable, $HVUR$. The outcome and control variables correspond to those in Table~\ref{table:shortlong}. Standard errors are clustered at the interacted labor market region (AMS-region) and year of labor market entry as well as calendar month level. Stars indicate significance levels:  *~$p<0.10$, **~$p<0.05$, ***~$p<0.01$.

\end{table} 

\end{landscape}

\end{document}